\documentclass[journal,onecolumn]{IEEEtran}
\usepackage{cite}
\usepackage{amsmath,amssymb,amsfonts}
\usepackage{algorithmic}
\usepackage{subfigure}
\usepackage{graphicx}
\usepackage{bbding}
\usepackage{textcomp}
\usepackage{tabularx}
\begin{document}
%
\title{Research on Sparsity Measures for Rotating Machinery Health Monitoring}
%
%
%

\author{Yudong~Cao,~\IEEEmembership{Graduate Student Member,~IEEE,}
        Minping~Jia,
        Jichao~Zhuang,
        and~Xiaoli~Zhao
\thanks{Yudong Cao, Jichao Zhuang and Prof. Minping Jia was with Southeast University, Southeast University, Nanjing,
Jiangsu, China.}
\thanks{Xiaoli Zhao was with Southeast University, Nanjing University of Science and Technology, Nanjing, Jiangsu, China.}
\thanks{Manuscript received XXX; revised XXX.}}

%
%



\maketitle

\begin{abstract}
Machine health management is one of the main research contents of PHM technology, which aims to monitor the health states of machines online and evaluate degradation stages through real-time sensor data. In recent years, classic sparsity measures such as kurtosis, Lp/Lq norm, pq-mean, smoothness index, negative entropy, and Gini index have been widely used to characterize the impulsivity of repetitive transients. Since smoothness index and negative entropy were proposed, the sparse properties have not been fully analyzed. The first work of this paper is to analyze six properties of smoothness index and negative entropy. In addition, this paper conducts a thorough investigation on multivariate power average function and finds that existing classical sparsity measures can be respectively reformulated as the ratio of multivariate power mean functions (MPMFs). Finally, a general paradigm of index design are proposed for the expansion of sparsity measures family, and several newly designed dimensionless health indexes are given as examples. Two different run-to-failure bearing datasets were used to analyze and validate the capabilities and advantages of the newly designed health indexes. Experimental results prove that the newly designed health indexes show good performance in terms of monotonic degradation description, first fault occurrence time determination and degradation state assessment.
\end{abstract}


%
\IEEEpeerreviewmaketitle

\section{Introduction}
%
%
%
%
\IEEEPARstart{T}{he} prognostics and Health Management (PHM) includes anomaly detection, fault diagnosis, and remaining useful life prediction\cite{Lei2018,Ma2021a,Lee2014a}. The research on PHM can improve the reliability and safety of mechanical equipment, which is related to production safety and economic benefits. Therefore, it is of great significance to study effective and accurate online health monitoring and degradation assessment methods for mechanical equipment, and then to carry out predictive management and maintenance. At the same time, with the maturity of industrial field monitoring technology, more and more sensor data can be collected to analyze the running state of mechanical equipment\cite{Zhang2021,Park2021a}. Health indexes (HIs) are designed to extract characteristic information from collected field data to identify and quantify historical and ongoing health processes of mechanical system components. The design of reasonable health indexes has become one of the main research directions of PHM technology, which has attracted more and more attention in recent years\cite{Miao2019}. Due to the complexity of mechanical systems and the benefit factors in actual production, it is impossible to frequently shut down the equipment to check the health status, and some early faults or damages are difficult to detect in time. Generally speaking, a ideal health index should have good monotonic performance and be able to detect abnormal states and initial fault occurrence time\cite{Baraldi2018b}.

Sparsity measures (SMs)\cite{Miao2020,Hou2021} as a kind of statistical parameters provide a new perspective for machine health index construction, including kurtosis,  $Lp/Lq$ norm, pq-mean, smoothness index, negative entropy, and Gini index, etc. In general, rotating machinery fault signals have two characteristics: one is impulsiveness and the other is cyclo-stationarity\cite{Li2016a}. In view of this, Dwyer and Antoni\cite{Dwyer1983,Antoni2006b,Antoni2006c} proposed spectral kurtosis successively, gave a formalization of spectral kurtosis to quantify the impulsiveness of repetitive transients, and realized its successful application in vibration-based condition monitoring. Wang et al.\cite{Wang2016} demonstrated the application of spectral kurtosis in fault detection and diagnosis of key components of rotating machinery and verified the ability of spectral kurtosis to quantify cyclo-stationarity. Zhong et al.\cite{Zhong2021b} proposed a weighted residual regression-based index to relieve the sensitivity of kurtosis and entropy to impulsive noise for gear and bearing degradation evaluation. The latest research mathematically proved that spectral kurtosis can be decomposed into squared L2/L1 norm and squared envelope\cite{Wang2018c}. The Gini index was first proposed in the economic field to measure the economic wealth inequality of residents of a country or region\cite{Dalton1920}. Later, it was introduced into the PHM field and considered a more effective criterion than kurtosis and correlated kurtosis for machine condition monitoring. Hurley et al.\cite{Hurley2009} compared several popular sparsity measures based on different attributes, including pq-mean and Gini index, etc. Miao et al.\cite{Miao2022} designed enhancing signal processing methods via Gini index and proposed two blind deconvolution methods based on Gini index and its variants.

In recent years, spectral negative entropy and smoothness index have been added to the sparsity measures family as new members. Antoni\cite{Antoni2016} introduced negative entropy in the field of thermodynamics as a sparse metric to characterize repetitive fault transients. Smoothness index\cite{Bozchalooi2007} was first designed to guide the selection of wavelet parameters, which can be expressed in mathematical form as the ratio of geometric mean and arithmetic mean. Miao et al.\cite{Miao2019a} applied singular value negentropy (SVN) to measure the periodicity of signal without prior knowledge. However, the sparse properties of negative entropy and smoothness index are rarely analyzed. This paper conducts an in-depth analysis of these two sparsity measures from six attributes, including Robin Hood, Scaling, Rising Tide, Cloning, Bill Gates, and Babies. In addition, this paper firstly investigates the multivariate power mean functions (MPMFs) and further finds that most of the currently used sparsity measures can be reformulated as ratios of different MPMFs. On this basis, a general paradigm are proposed for the design of new sparsity measures for machine health monitoring. The main contributions of this paper can be summarized as follows:

(1)The sparse properties of negative entropy and smoothness index are fully studied from six aspects for the first time, and the detailed proofs and final conclusions are given.

(2) The ratio of multivariate power average function is introduced into the domain of online machine monitoring. Mathematically, six existing sparsity measures (kurtosis, $Lp/Lq$ norm, pq-mean, smoothness index, negative entropy, and Gini index) are respectively reformulated into the form of ratios of multivariate power mean functions.

(3) A general paradigm of health index design based on the ratios of various MPMFs are proposed, and some examples of dimensionless health indexes designed from different perspectives are presented. The capabilities and advantages of newly designed HIs are specifically analyzed and validated by two run-to-failure bearing datasets. 

The rest of the paper is organized as follows. Section II analyzes the sparse properties of smoothness index and negative entropy. Six existing sparsity measures are reformulated in the form of ratios of different MPMFs and a general paradigm are introduced for the expansion of sparsity measures family in Section III. In Section IV, two run-to-failure bearing datasets are employed to validate the generalization and reproducibility of proposed paradigm and newly designed health indexes. Finally, conclusions are drawn in Section V.

\section{Sparse Attribute Analysis of Smoothness Index and Negative Entropy}
\subsection{Six Attributes of Ideal Sparsity Measures}
Suppose a sparse measure $S$ as a function with the mapping as follows
\begin{equation}
S:\left(\bigcup_{n \geq 1} \mathbb{X}^{n}\right) \rightarrow \mathbb{R}
\end{equation} where $\mathbb{X}^{n}$ represents a complex vector of length N. Metric $S$ maps complex vectors into a real number. According to previous research and inspiration from economics\cite{Dykstra1988,Hurley2009}, it is generally believed that ideal SMs should own six attributes.

\textbf{D1. Robin Hood}

Dalton's 1st Law states that Robin Hood decreases sparsity. Robin Hood operation subtracts a positive constant $\alpha$ from any relatively large element $x_i$ in the vector, and adds $\alpha$ to a relatively small element $x_j$. After operation, new $x_i$ is still larger than new $x_j$. The mapping value of the sparse metric $S$ on the new vector becomes smaller. Robin Hood can be formally defined as follows:
\begin{equation}
S \left ( \left [x_1 \dots x_i-\alpha \dots x_j+\alpha \dots \right ] \right )<S(\overset\rightarrow{x})
	\label{eq:robin_hood}
\end{equation}
for all $\alpha$, $x_i$, $x_j$ such that $x_i>x_j$ and $0<\alpha<\frac{x_i-x_j}{2}$.

\textbf{D2. Scaling}

Dalton's modified 2nd Law states that Sparsity is scale invariant. After all elements in the vector are multiplied by a constant factor $\alpha$, the sparse value of the vector remains invariant. Scaling can be formally defined as follows:
\begin{equation}
S(\alpha \overset\rightarrow{x})=S(\overset\rightarrow{x}),\forall \alpha \in \mathbb{R}, \alpha>0
\end{equation}

\textbf{D3. Rising Tide}

Dalton's 3rd Law states that adding a constant to each element in vector can reduce sparsity. After adding a positive constant $\alpha$ to all elements in the vector, the sparse value of the vector becomes smaller. We exclude the case where all elements are equal as this is equivalent to \textbf{D2}. Rising Tide can be formally defined as follows:
\begin{equation}
S(\alpha+\overset\rightarrow{x})<S(\overset\rightarrow{x}),\forall \alpha \in \mathbb{R}, \alpha>0
\end{equation}

\textbf{D4. Cloning}

Dalton's 4th Law states that sparsity is invariant under cloning operation. The sparse value is unchanged after the vector is joined with itself. Cloning can be formally defined as follows:
\begin{equation}
S(\vec{x})=S(\vec{x} \| \vec{x})=S(\vec{x}\|\vec{x}\| \vec{x})=S(\vec{x}\|\vec{x}\| \cdots \| \vec{x})
\end{equation}

\textbf{P1. Bill Gates}

An element $x_i$ in the vector becomes infinitely large, and the sparse value of the vector also tends to infinitely large. Bill Gates can be formally defined as follows:

$\forall i$, $\exists \beta=\beta_{i}>0$, such that $\forall \alpha>0$
\begin{equation}
	S \left ( \left [x_1 \dots x_i+\beta+\alpha \dots  \right ] \right )>S\left ( \left [x_1 \dots x_i+\beta \dots \right ] \right )
\end{equation}

\textbf{P2. Babies}

After adding an element of zero into the vector, and the sparse value of the new vector will increase. Babies can be formally defined as follows:
\begin{equation}
S(\vec{x} \| 0)>S(\vec{x})
\end{equation}

\subsection{Attributes of SI and NE}
In previous work, Hurely\cite{Hurley2009} proved the sparsity of kurtosis, Hoyer, Pq-mean and Gini index and other classical sparse measures. Since smoothness index and negative entropy were added to the sparsity measures family, their sparse properties have not been fully studied. The first work of this paper is to analyze the sparse quantification capability of smoothness index and negative entropy from the above six attributes. The result of mathematical derivation shows that smoothness index satisfies \textbf{nonnegativity}, \textbf{D2} and \textbf{D4}; and negative entropy satisfies \textbf{nonnegativity}, \textbf{D1}, \textbf{D2}, \textbf{D3}, \textbf{D4}, \textbf{P1}, as shown in Table \ref{attribute}. The detailed proofs are given in Appendix. Based on our research with Dr. Hurely, the following conclusion can be drawn: Pq-mean, Hoyer, Gini index and negative entropy have better sparse quantification capability among existing popular sparse measures.

\begin{table}[!h]
	\centering
	\caption{Detailed Attributes of SI and NE.}
	\label{attribute}
	\begin{tabularx}{\columnwidth}{X>{\centering}p{44pt}>{\centering}p{8pt}>{\centering}p{8pt}>{\centering}p{8pt}>{\centering}p{8pt}>{\centering}p{8pt}p{8pt}}
		\hline
		\hline
		\textbf{Sparsity Measures}     & \textbf{Nonnegativity}    & \textbf{D1} & \textbf{D2} & \textbf{D3} & \textbf{D4} & \textbf{P1} & \textbf{P2} \\
		\hline
		Smoothness index   & \Checkmark  & \XSolid   &  \Checkmark  & \XSolid  & \Checkmark & \XSolid &  \XSolid      \\
		Negative entropy  & \Checkmark  & \Checkmark  & \Checkmark & \Checkmark  & \Checkmark & \XSolid  & \Checkmark      \\
		\hline
		\hline
	\end{tabularx}
	\vspace{-0.25cm}
\end{table}

\section{Reformulation and Family Extension of Sparsity Measures}

\subsection{Definition of Multivariate Power Mean Function}
Suppose a function has the form of Eq. (\ref{eq:MPMF1}), then it can be called a power mean function, where $a_i(i=1,2,\dots,n)$ are incomplete equal positive constants. This function is defined on the entire real number axis, and there is always $F(x)>0$. Obviously, no matter how x varies, the power mean function does not tend to infinity. Therefore, power mean function is bounded. If $min{\left\{a_i\right\}}=a$, $max{\left\{a_i\right\}}=A$, then $a\le F(x)\le A$.

\begin{equation}
F(x)= \begin{cases}\min \left(a_{1}, \ldots, a_{n}\right) & \text { if } x=-\infty \\ \left(\frac{a_{1}^{x}+\cdots+a_{n}^{x}}{n}\right)^{1 / x} & \text { if } x \in \mathbb{R} \backslash\{0\} \\ \sqrt[n]{a_{1} \cdots a_{n}} & \text { if } x=0 \\ \max \left(a_{1}, \ldots, a_{n}\right) & \text { if } x=+\infty\end{cases}\label{eq:MPMF1}
\end{equation}

Based on the above, we define a function with the form $\Gamma(\mathbf{x}, y)=[\frac{1}{N} \sum_{n=1}^{N}\left(x_{n}\right)^{y}]^{\frac{1}{y}}$ as a multivariate power mean function (MPMF), where $\mathbf{x}=\left[x_{1}, x_{2}, \cdots x_{n}\right]$ represents a row vector. The classical RMS metric can be expressed as $\Gamma(\mathbf{x}, 2)$ in the form of MPMF. In addition, the multivariate power mean inequality needs to be introduced: If $y_1>y_2$, then $\Gamma\left(\mathbf{x}, y_{1}\right) \geq \Gamma\left(\mathbf{x}, y_{2}\right)$, and the equality holds if and only if $x_1=x_2=\cdots x_n$. The proof only needs to use Jensen inequality in mathematical analysis (i.e., up and down convexity), and take the auxiliary function $f(x)=x^{a}$.
\vspace{-3mm}

\subsection{Reformulation of Six Classic Sparsity Measures}

Assume that a raw vibration signal is denoted as $x[n],n=1,2,3,\dots N$. Then, a bandpass filtered signal $x_{l,h}[n]$ can be obtained by a using bandpass filter with a passband $[l,h]$. Further, a complex signal $\overline{x_{l, h}}[n]=x_{l, h}[n]+j \cdot \text {\textit{Hilbert}}\left\{x_{l, h}[n]\right\}$ and square envelope $S E_{l, h}[n]=\left|\overline{x_{l,h}}[n]\right|^{2}$ can be obtained by Hilbert transform. To simplify symbols, $S E_{l, h}[n]$ is replaced by $\overline{x_{l, h}}[n]$ for the rest of the paper.

Six classic sparsity measures are respectively reformulated as the ratios of different multivariate power mean functions in this section, including spectral kurtosis, spectral $Lp/Lq$ norm, pq-mean, spectral smoothness index, spectral negative entropy and Gini index. The following subsections give the details of the reformulation. 

\subsubsection{Spectral Kurtosis}

The original definition of kurtosis is the normalized 4th order center moment. Spectral kurtosis can be reformulated by combining the ratios of MPMFs as follows:
\vspace{-1mm}
\begin{equation}
	\begin{aligned}
		SK &=\frac{m_{4}\left\{\left|\overline{x_{l,h}}[n]\right|\right\}}{\left(m_{2}\left\{\left|\overline{x_{l,h}}[n]\right|\right\}\right)^{2}}
		=\frac{\sum_{n=1}^{N} \frac{S E_{l, h}^{2}[n]}{N}}{\left(\sum_{n=1}^{N} \frac{S E_{l, h}[n]}{N}\right)^{2}}\\
		&=\left(\frac{\sqrt{\sum_{n=1}^{N} X_{l, h}^{2}[n] / N}}{\sum_{n=1}^{N} X_{l, h}[n] / N}\right)^{2}=\left[\frac{\Gamma(\mathbf{X}, 2)}{\Gamma(\mathbf{X}, 1)}\right]^{2}
	\end{aligned}
\end{equation}
\vspace{-3mm}
\subsubsection{Spectral $Lp/Lq$ norm}

Spectral $Lp/Lq$ norm can be reformulated as the ratio of two different MPMFs as follows:
\\when $p>q>0$
\vspace{-3mm}
\begin{equation}
	\begin{aligned}
		GSK&=N^{\frac{1}{q}-\frac{1}{p}} \frac{\left\|S E_{l, h}[n]\right\|_{L_{P}}}{\left\|S E_{l, h}[n]\right\|_{L q}}-\frac{\sqrt[p]{p !}}{\sqrt[q]{q !}}\\&=\frac{\left(\sum_{n=1}^{N} \frac{X_{l, h}^{p}[n]}{N}\right)^{\frac{1}{p}}}{\left(\sum_{n=1}^{N} \frac{X_{l, h}^{q}[n]}{N}\right)^{\frac{1}{q}}}-\frac{\sqrt[p]{p !}}{\sqrt[q]{q !}}=\frac{\Gamma(\mathbf{X}, p)}{\Gamma(\mathbf{X}, q)}-\frac{\sqrt[p]{p !}}{\sqrt[q]{q !}}
	\end{aligned}
\end{equation}
when $p>q=0$
\vspace{-3mm}
\begin{equation}
	\begin{aligned}
		G S K&=N^{-\frac{1}{p}} \frac{\left\|S E_{l, h}[n]\right\|_{L_{P}}}{\sqrt[N]{\prod_{n=1}^{N} S E_{l, h}[n]}}-\frac{\sqrt[p]{p !}}{e^{-\gamma}}\\&=\frac{\left(\sum_{n=1}^{N} X_{l, h}^{p}[n] / N\right)^{\frac{1}{p}}}{\sqrt[N]{\prod_{n=1}^{N} X_{l, h}[n]}}-\frac{\sqrt[p]{p !}}{e^{-\gamma}}=\frac{\Gamma(\mathbf{X}, p)}{e^{\Gamma(\ln (\mathbf{X}), 1)}}-\frac{\sqrt[p]{p !}}{e^{-\gamma}}
	\end{aligned}
\end{equation}
where $e^{\Gamma(\ln (\mathbf{X}), 1)}$ is equal to $\Gamma(\mathbf{X},0)$; $\gamma$ is the Euler-Mascheroni constant 0.5772156649….

\subsubsection{$Pq-mean$}

Pq-mean is also a widely used sparsity measure with good performance. By the ratio of multivariate power mean functions, the pq-mean can be reformulated as follows:

when $0<p\le1,q>1$
\begin{equation}
	\begin{aligned}
		pq-\text { mean }&=-\left(\frac{1}{N} \sum_{n=1}^{N} X_{l, h}^{p}[n]\right)^{\frac{1}{p}}\left(\frac{1}{N} \sum_{n=1}^{N} X_{l, h}^{q}[n]\right)^{-\frac{1}{q}}\\&=-\frac{\Gamma(\mathbf{X}, p)}{\Gamma(\mathbf{X}, q)}
	\end{aligned}
\end{equation}

\subsubsection{Spectral smoothness}
Spectral smoothness index can also be reformulated in the form of the ratio of different MPMFs as follows:
\begin{equation}
	\begin{aligned}
		SI&=\frac{\sqrt[N]{\prod_{n=1}^{N}S E_{l, h}[n]}}{\frac{\sum_{n=1}^{N} S E_{l, h}[n]}{N}}=\frac{\sqrt[N]{\prod_{n=1}^{N}X_{l, h}[n]}}{\frac{\sum_{n=1}^{N} X_{l, h}[n]}{N}}\\&=\frac{e^{\Gamma(\ln (\mathbf{X}), 1)}}{\Gamma(\mathbf{X}, 1)}
	\end{aligned}
\end{equation}

\subsubsection{Spectral negative entropy}

Similarly, spectral negative entropy can be redefined as follows:
\begin{equation}
	\begin{aligned}
		SNE&=\left\langle\frac{\left|\overline{x_{l, h}}[n]\right|^{2}}{\left\langle\left|\overline{x_{l, h}}[n]\right|^{2}\right\rangle} \ln \left(\frac{\left|\overline{x_{l, h}}[n]\right|^{2}}{\left\langle\left|\overline{x_{l,h}}[n]\right|^{2}\right\rangle}\right)\right\rangle\\&=\left\langle\frac{S E_{l, h}[n]}{\left\langle S E_{l, h}[n]\right\rangle} \ln \left(\frac{S E_{l, h}[n]}{\left\langle S E_{l, h}[n]\right\rangle}\right)\right\rangle\\&=\left\langle\frac{X_{l, h}[n]}{\Gamma(\mathbf{X}, 1)} \ln \left(\frac{X_{l, h}[n]}{\Gamma(\mathbf{X}, 1)}\right)\right\rangle\\&=\frac{\Gamma\left(\frac{\mathbf{X}}{\Gamma(\mathbf{X}, 1)} \ln \left(\frac{\mathbf{X}}{\Gamma(\mathbf{X}, 1)}\right), 1\right)}{\Gamma(1,1)}
	\end{aligned}
\end{equation}

By further mathematical derivation, the following relationship between SNE and SI can be found:
\begin{equation}
	\begin{aligned}
		S N E&=\frac{\left(\sum_{n=1}^{N} X_{l, h}[n] \frac{1}{N} \ln \left(\frac{X_{l, h}[n]}{\left\langle X_{l, h}[n]\right\rangle}\right)\right)}{\ln \left(\frac{\sqrt[N]{\prod_{n=1}^{N}X_{l, h}[n]}}{\sum_{n=1}^{N} X_{l, h}[n] / N}\right)\left(\sum_{n=1}^{N} X_{l, h}[n] / N\right)} \ln (S I)\\&
		=\frac{\left(\frac{1}{N} \sum_{n=1}^{N} X_{l, h}[n] \ln \left(\frac{X_{l, h}[n]}{\Gamma(\mathbf{x}, 1)}\right)\right)}{\ln \left(\frac{e^{\Gamma(\ln (\mathbf{X}), 1)}}{\Gamma(\mathbf{X}, 1)}\right)\Gamma(\mathbf{X}, 1)} \ln (S I)\\&=\frac{\Gamma\left(\mathbf{X} \ln \left(\frac{\mathbf{X}}{\Gamma(\mathbf{X}, 1)}\right), 1\right)}{\Gamma(\mathbf{X}, 1)(\Gamma(\ln (\mathbf{X}), 1)-\ln (\Gamma(\mathbf{X}, 1)))} \ln (S I)
	\end{aligned}
\end{equation}

\subsubsection{Gini index}

The Gini index was originally proposed in the field of economics. It was later introduced into the field of fault diagnosis and has become a reliable measure in signal processing techniques. The value of the Gini index ranges from 0 to 1. The definition of Gini index is shown in Eq. (\ref{eq:Gini}).
\begin{equation}
	G I=1-2 \sum_{n=1}^{N} \frac{X^{\text {order }}[n]}{\|X[n]\|_{L 1}}\left(\frac{N-n+0.5}{N}\right)\label{eq:Gini}
\end{equation}
Gini index can be reformulated as the ratio of two different MPMFs as follows:
\begin{equation}
	\begin{aligned}
		G I&=1-2 \sum_{n=1}^{N}\left(\left(\frac{N-n+\frac{1}{2}}{N}\right) \times \frac{S E_{l, h}^{\text {order }}[n]}{\left\|S E_{l, h}[n]\right\|_{L 1}}\right)\\&=1-2 \sum_{n=1}^{N}\left(\left(\frac{N-n+\frac{1}{2}}{N^{2}}\right) \times \frac{X_{l, h}^{\text {order }}[n]}{\sum_{n=1}^{N} X_{l, h}[n] / N}\right)\\
		&=1-\frac{\sum_{n=1}^{N}\left(2\left(\frac{N-n+\frac{1}{2}}{N^{2}}\right) \times X_{l, h}^{\text {order }}[n]\right)}{\sum_{n=1}^{N} X_{l, h}[n] / N}\\&=1-\frac{\Gamma\left(2\left(\frac{N-n+\frac{1}{2}}{N}\right) \times \mathbf{X}_{\text {order }}, 1\right)}{\Gamma(\mathbf{X}, 1)}
	\end{aligned}
\end{equation}
where $2\left(\frac{N-n+\frac{1}{2}}{N}\right) \times \mathbf{X}_{\text {order}}$ can be viewed as a weighted and ordered vector.

An important point needs to be emphasized additionally. Hou et al.\cite{Hou2021a} expressed the classical sparsity measures as ratio of quasi-arithmetic means in their previous work. The multivariate power mean we mentioned can generally be expressed as a kind of generalized arithmetic mean. Through multivariate power mean function, the reformulation expression can be made more popular. At the same time, it is more convenient to reconstruct the spectral negative entropy and Gini index.

\subsection{Family Extension of Sparsity Measures}

Inspired by reformulation of the six classical sparsity measures in Subection B, we propose a general paradigm based on ratios of multivariate power mean functions for designing more advanced health indexes for machine health monitoring and degradation state assessment.

Complex health indexes designed by combining ratios of the sum of multiple MPMFs:
\begin{equation}
	P H I=\lambda \frac{\sum_{m=1}^{M} p_{m} \Gamma\left(\mathbf{x}, a_{m}\right)}{\sum_{n=1}^{N} q_{n} \Gamma\left(\mathbf{x}, b_{m}\right)}+c\label{eq:PHI}
\end{equation}

Complex health indexes designed by combining ratios of the product of multiple MPMFs:

\begin{equation}
	M H I=\lambda \frac{\prod_{m=1}^{M} p_{m} \Gamma\left(\mathbf{x}, a_{m}\right)}{\prod_{n=1}^{N} q_{n} \Gamma\left(\mathbf{x}, b_{m}\right)}+c\label{eq:MHI}
\end{equation}
where $p_m$ and $q_n$ represent different weights of the power average function respectively, and satisfy $\sum_{m=1}^{M} p_{m}=1, \sum_{n=1}^{N} q_{n}=1$, $\lambda$ is a tradeoff parameter, and $c$ is an offset coefficient providing a baseline for indexes. From previous studies, it is clear that $\Gamma(\alpha\mathbf{x}, y)=\alpha\Gamma(\mathbf{x}, y)$ and $\Gamma(\mathbf{x}\|\mathbf{x}\| \cdots \| \mathbf{x},y)=\Gamma(\mathbf{x}, y)$ , so $PHI$ and $MHI$ satisfy at least two sparse attributes: \textbf{D2} and \textbf{D4}.

According to Eq. (\ref{eq:PHI}) and Eq. (\ref{eq:MHI}), we can give a general paradigm as follows:

\begin{equation}
	GHI=\sum_{j} PHI+\sum_{j} MHI
\end{equation}

The specific expressions for the four newly designed HIs as examples are as follows:
\begin{equation}
	\begin{aligned}
HI_{1}=1-\frac{\Gamma(\mathbf{X}+\tau,-2)+\Gamma(\mathbf{X}+\tau,-1)}{\Gamma(\mathbf{X}+\tau,-1)+\Gamma(\mathbf{X}+\tau, 1)}
	\end{aligned}
\end{equation}

\begin{equation}
	H I_{2}=1-\frac{\Gamma(\mathbf{X}+\tau,-2)+\Gamma(\mathbf{X}+\tau,-2)}{e^{\Gamma(\ln (\mathbf{X}+\tau), 1)}+\Gamma(\mathbf{X}+\tau,-1)}
\end{equation}

\begin{equation}
	H I_{3}=1-\frac{\Gamma(\mathbf{X}+\tau,-2) \Gamma(\mathbf{X}+\tau,-2)}{e^{\Gamma(\ln (\mathbf{X}+\tau), 1)} \Gamma(\mathbf{X}+\tau,-1)}
\end{equation}

\begin{equation}
	H I_{4}=1-\frac{e^{\Gamma(\ln (\mathbf{X}+\tau), 1)} \Gamma(\mathbf{X}+\tau,-2)}{\Gamma(\mathbf{X}+\tau, 1) \Gamma(\mathbf{X}+\tau,-1)}
\end{equation}

where $\tau$ represents slip coefficient which is related to shifted percentage \cite{Hou2021b}. In our case, $\tau$ is set to 0.065. At the same time, an extra property of the newly designed HIs need to be emphasized. According to multivariate power mean inequality, $HI_1$-$HI_4$ have strict upper and lower bounds $[0,1)$. Therefore, it is helpful to set a reasonable and effective failure threshold for machine health monitoring under different operating conditions.

\section{Validation And Analysis Based On Two Run-To-Failure Bearing Datasets }

\subsection{Case Study I: IMS Bearing Experimental Dataset}

\begin{figure}[!t]
	\centerline{\includegraphics[width=\columnwidth]{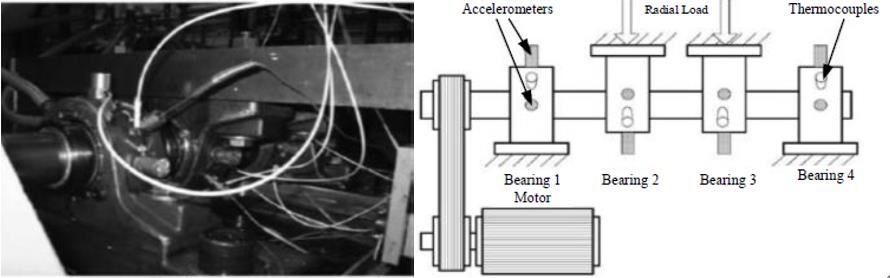}}
	\caption{The bearing experimental platform of IMS.}
	\label{IMS}
\end{figure}

The accelerated bearing degradation experimental dataset provided by the NSF I/UCR Center for Intelligent Maintenance Systems was firstly used to verify the effectiveness and superiority of the designed health indexes. The experimental platform is shown in Fig. \ref{IMS} with four bearings installed on one shaft. The sampling frequency is set at 20KHz, and 20,480 points are collected each time with an interval of 1 second. Set No.2 collected a total of 984 files in Channel 1. The bearing finally produced an outer ring failure, and the outer ring failure frequency could be calculated as 236.4Hz according to the theoretical formula of failure frequency.

\begin{figure}[!t]
	\centering
	\subfigure{
		\begin{minipage}[t]{0.5\linewidth}
			\centering
			\includegraphics[width=9cm,height=6.74cm]{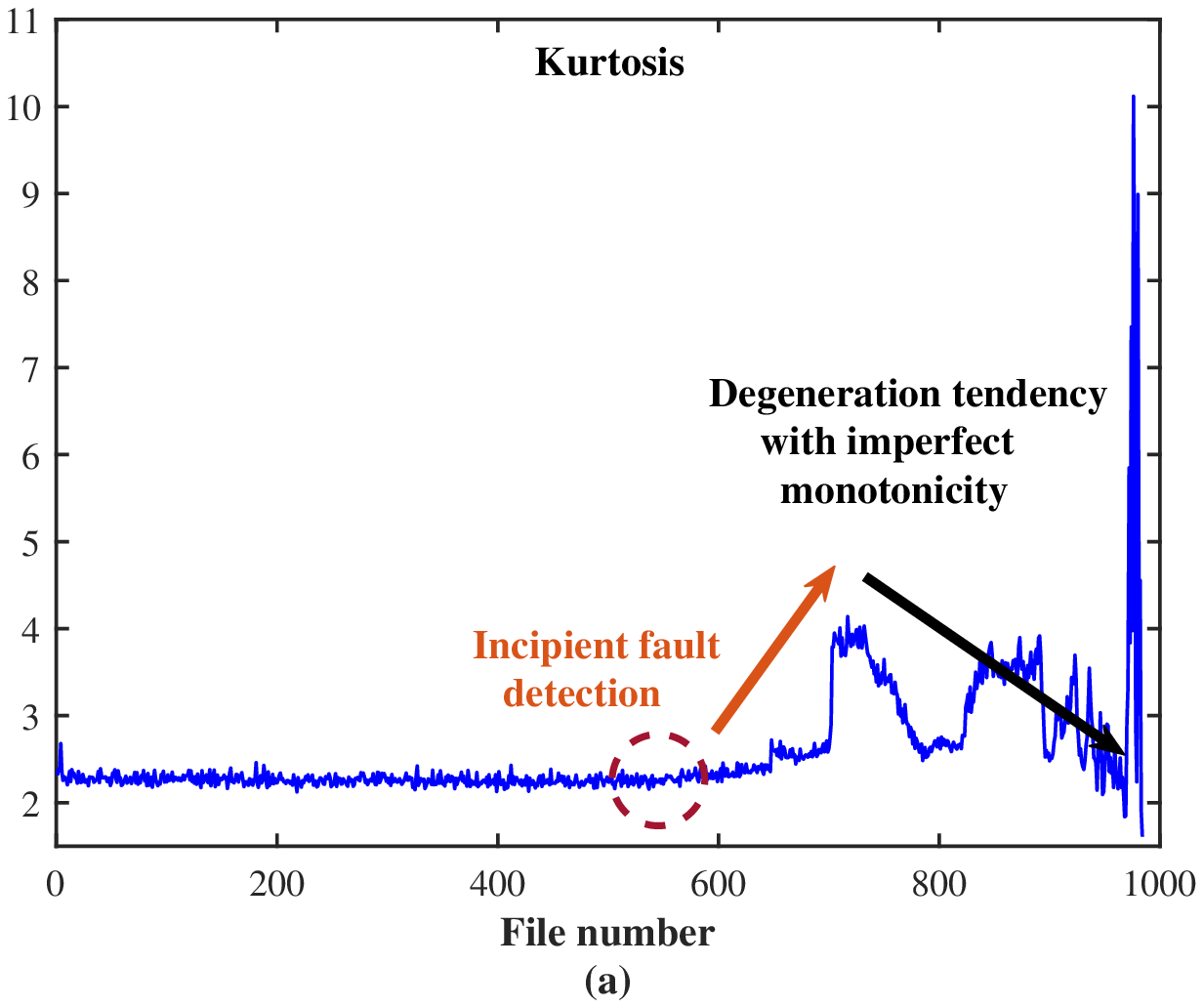}
		\end{minipage}%
	}%
	\subfigure{
		\begin{minipage}[t]{0.5\linewidth}
			\centering
			\includegraphics[width=9cm,height=6.74cm]{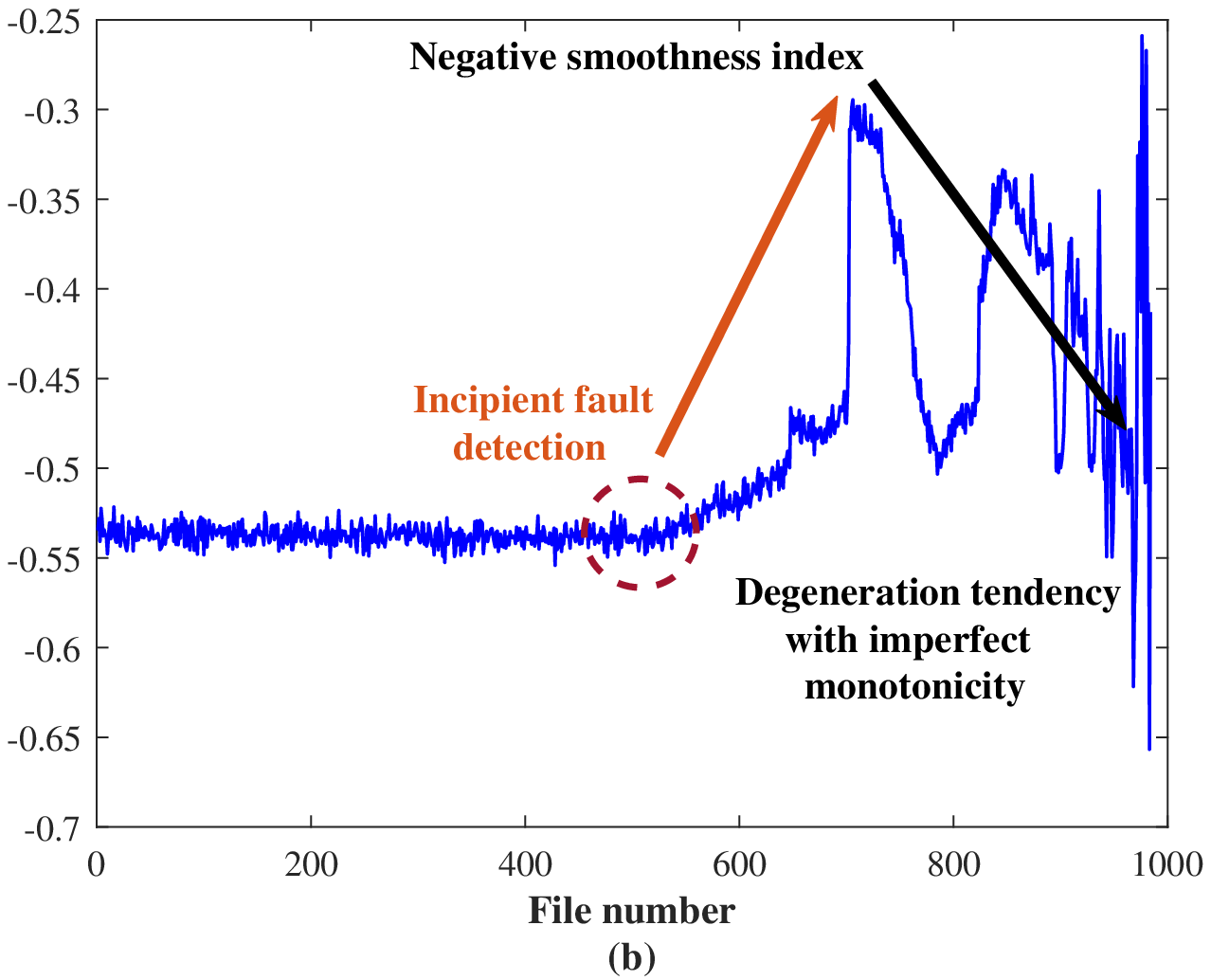}
		\end{minipage}%
	}%

	\subfigure{
		\begin{minipage}[t]{0.5\linewidth}
			\centering
			\includegraphics[width=9cm,height=6.74cm]{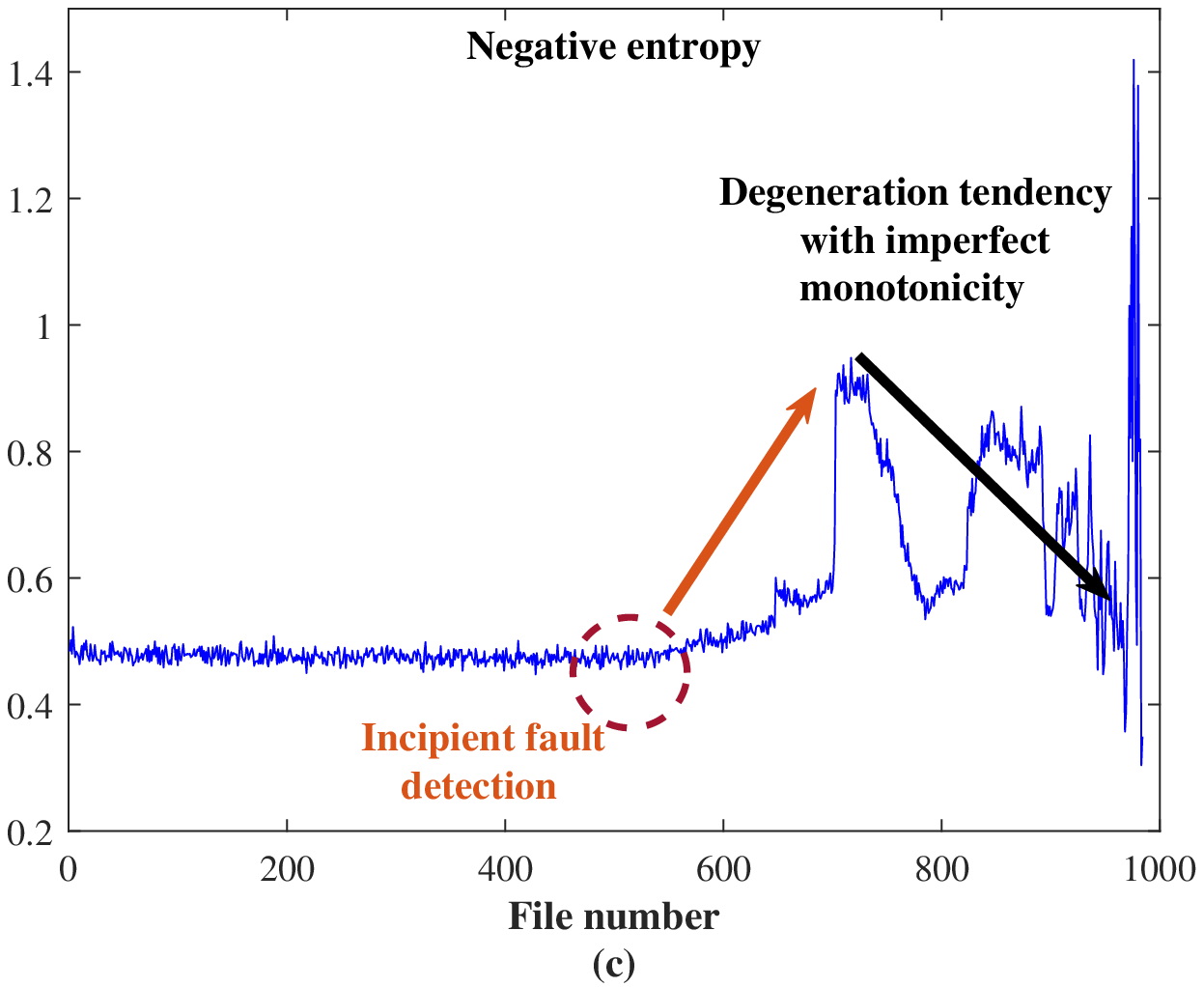}
		\end{minipage}%
	}%
	\subfigure{
		\begin{minipage}[t]{0.5\linewidth}
			\centering
			\includegraphics[width=9cm,height=6.74cm]{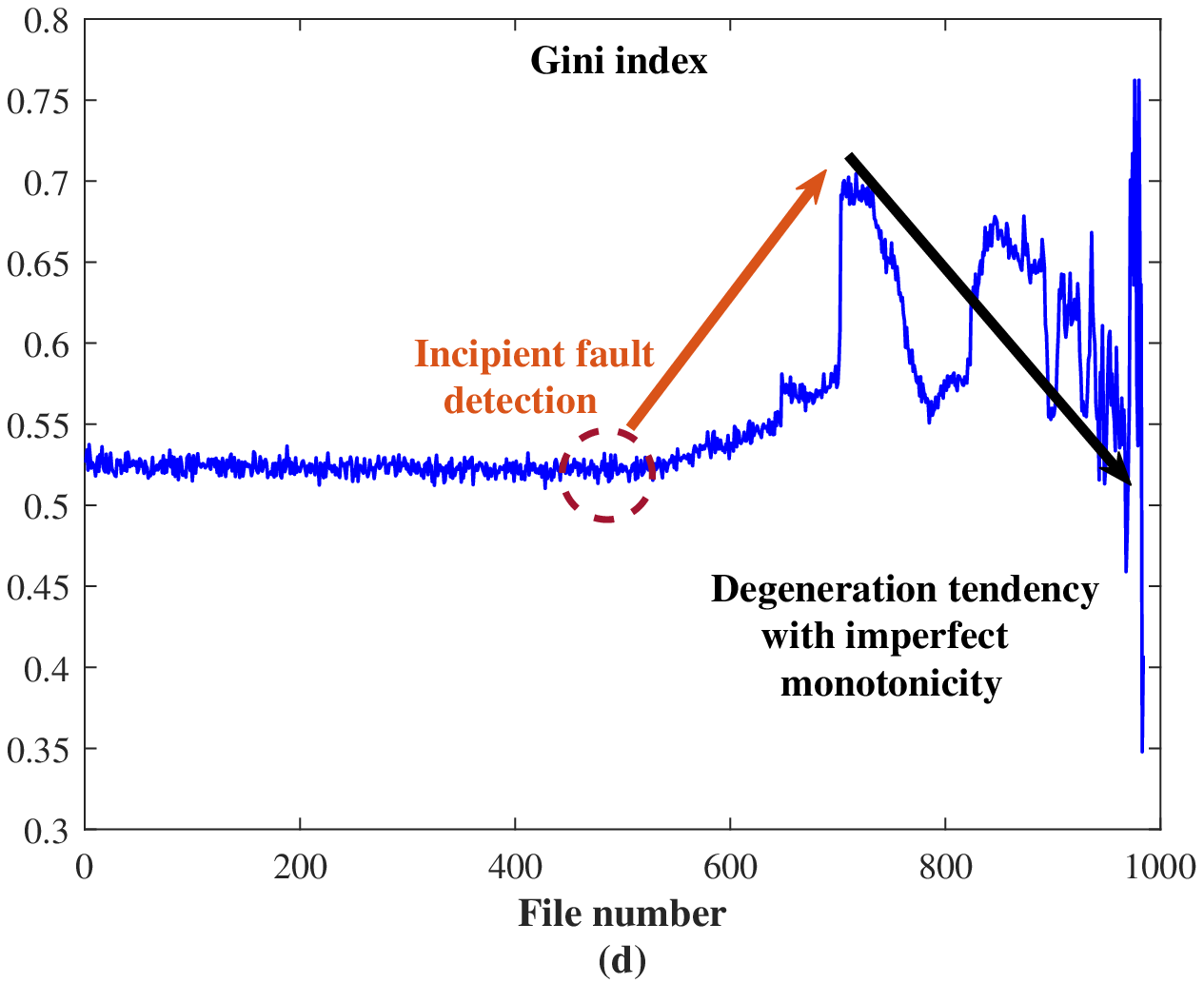}
		\end{minipage}%
	}%
	
	\centering
	\caption{Degradation curves plotted by classic sparsity measures: (a) spectral kurtosis; (b) spectral smoothness index; (c) spectral negative entropy; (d) Gini index.}\label{classic IMS}
\end{figure}

\begin{figure}[t]
	\centering
	\subfigure{
		
		\includegraphics[width=15cm,height=5cm]{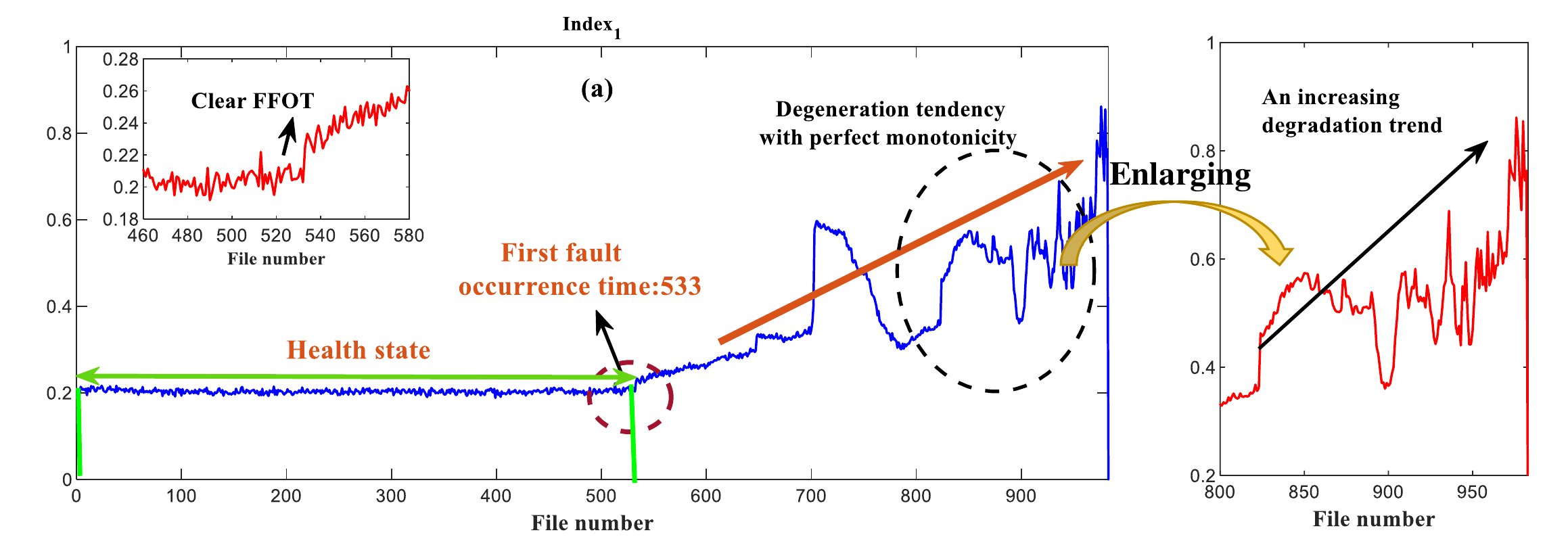}
		
	}%

	\subfigure{
		
		\includegraphics[width=15cm,height=5cm]{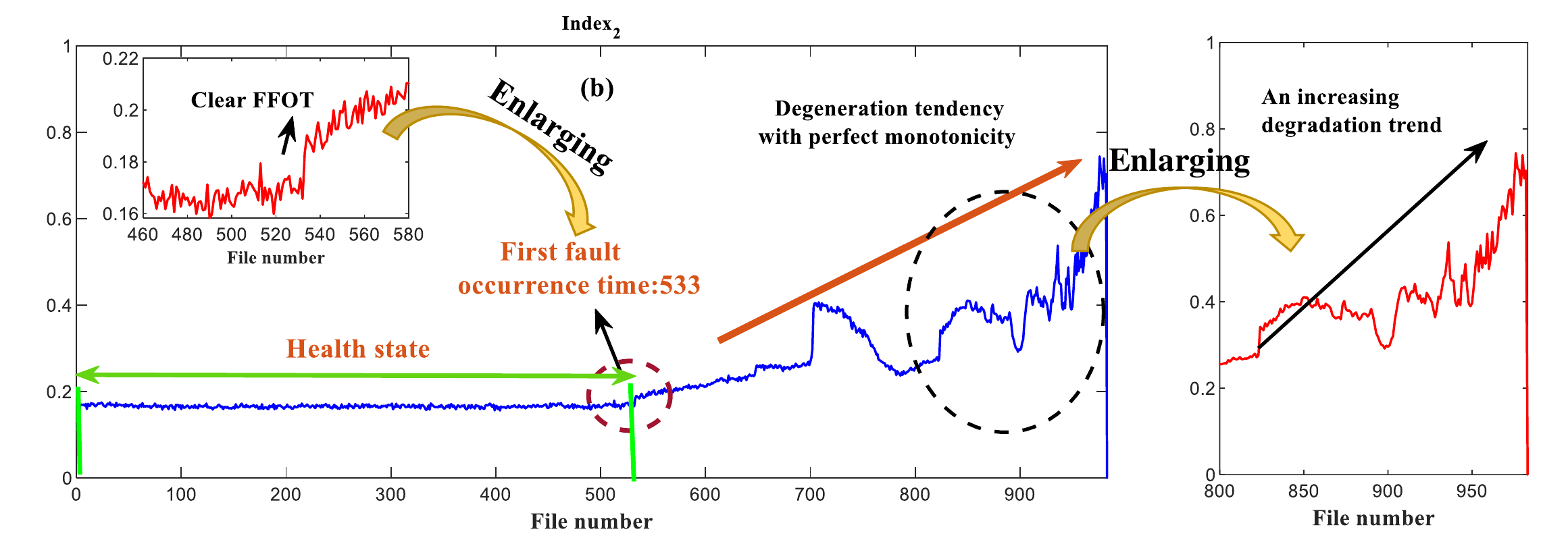}
		
	}%

	\subfigure{
		
		\includegraphics[width=15cm,height=5cm]{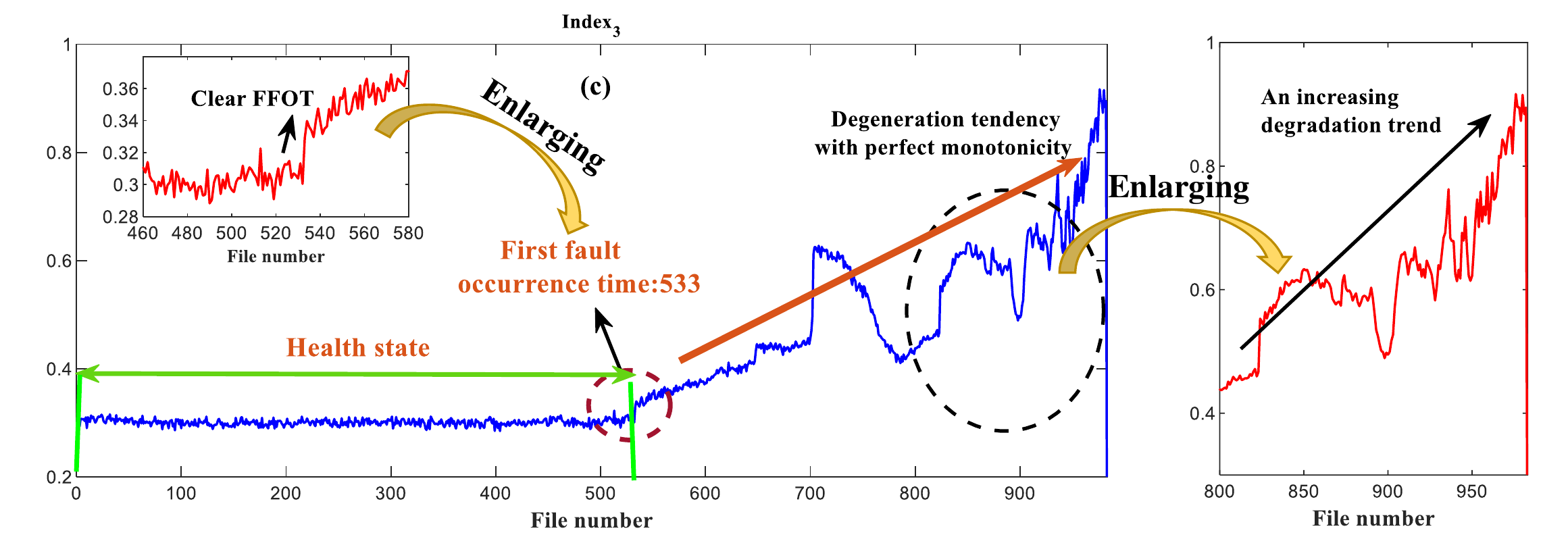}
		
	}%

	\subfigure{
		\includegraphics[width=15cm,height=5cm]{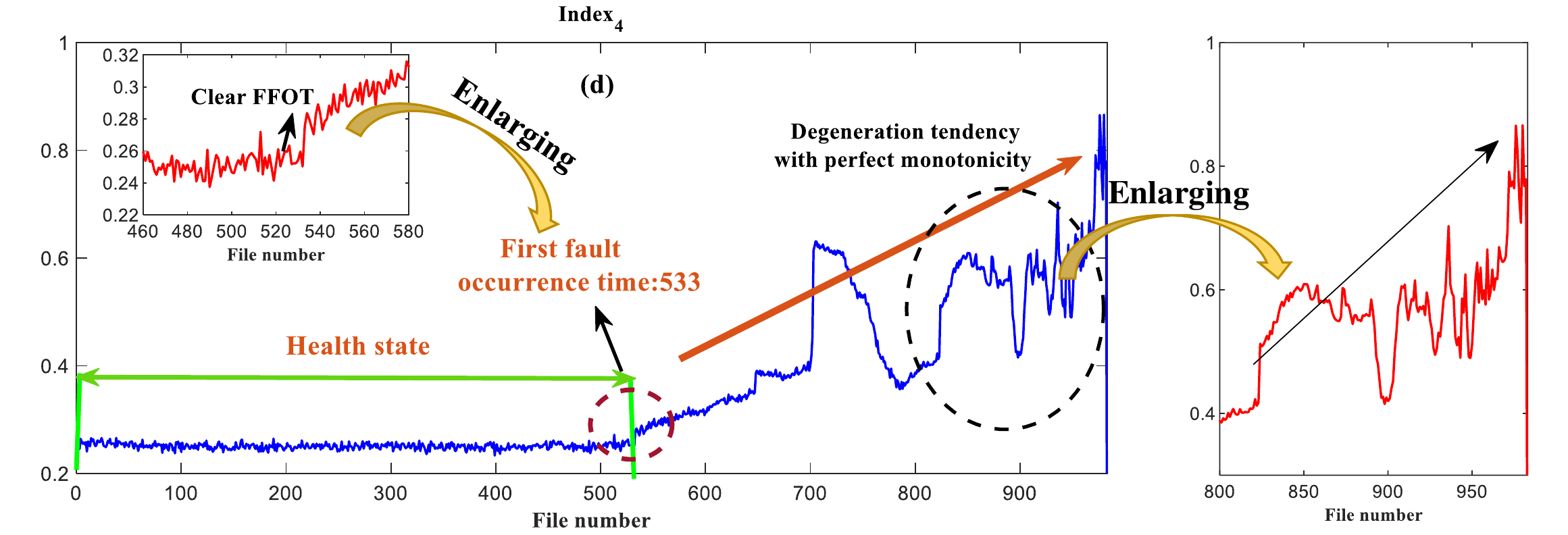}
		
	}%
	
	\centering
	\caption{Bearing degradation curves plotted by (a) Index1; (b) Index2; (c) Index3; (d) Index4.}\label{index IMS}
\end{figure}

The bearing degradation curves are plotted by classic spectral kurtosis, spectral smoothness index, spectral negative entropy and Gini index as shown in Fig. \ref{classic IMS}, where the real values of spectral smoothness index are multiplied by -1 to achieve an increasing trend. From the figures, it can be seen that these classical sparsity measures can reflect the bearing degradation trend to a certain extent and help in incipient fault detection, but they cannot characterize the bearing degradation monotonically and the degradation curve is highly fluctuant.
Our designed index degradation curves are shown in Fig. \ref{index IMS}. As can be seen from the figures, $HI_1$-$HI_4$ have good monotonic degradation trends with less fluctuation. Enlarged pictures around FFOT at file No.460 to No.580 quantified by designed  indexes are respectively plotted tt the top left of each figure.
It can be easily concluded that $HI_1$-$HI_4$ can detect the first failure occurrence time (FFOT) more precisely. In the following subsections, we will further evaluate the performance of the designed health metrics in three aspects.

\subsubsection{First Fault Occurrence Time (FFOT) Determination and Threshold Analysis}\

\begin{figure}[t]
	\centering
	\subfigure{
		
		\includegraphics[width=15cm,height=5cm]{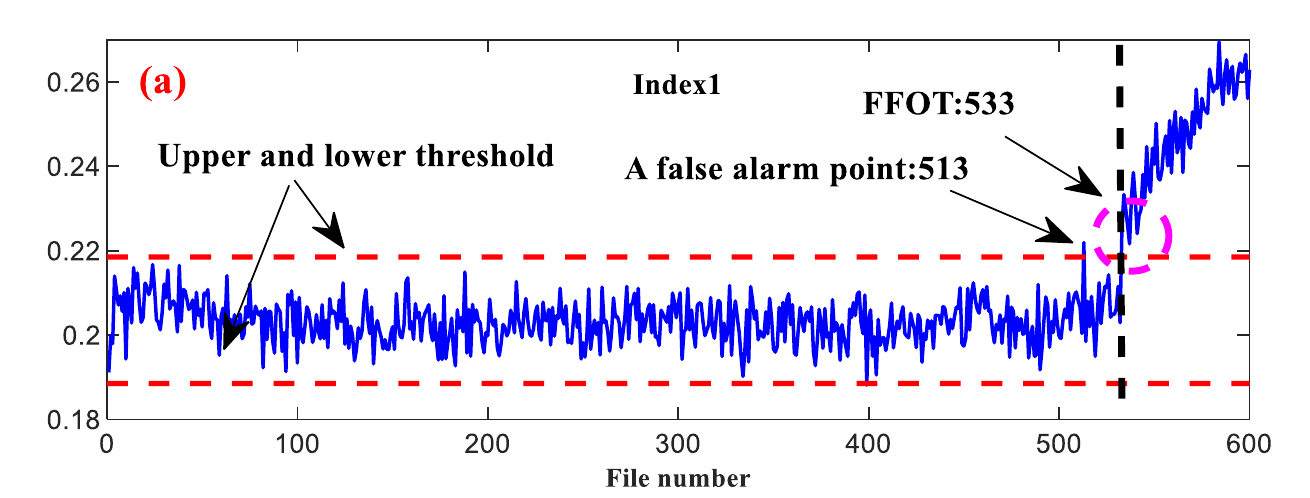}
		
	}%

	\subfigure{
		
		\includegraphics[width=15cm,height=5cm]{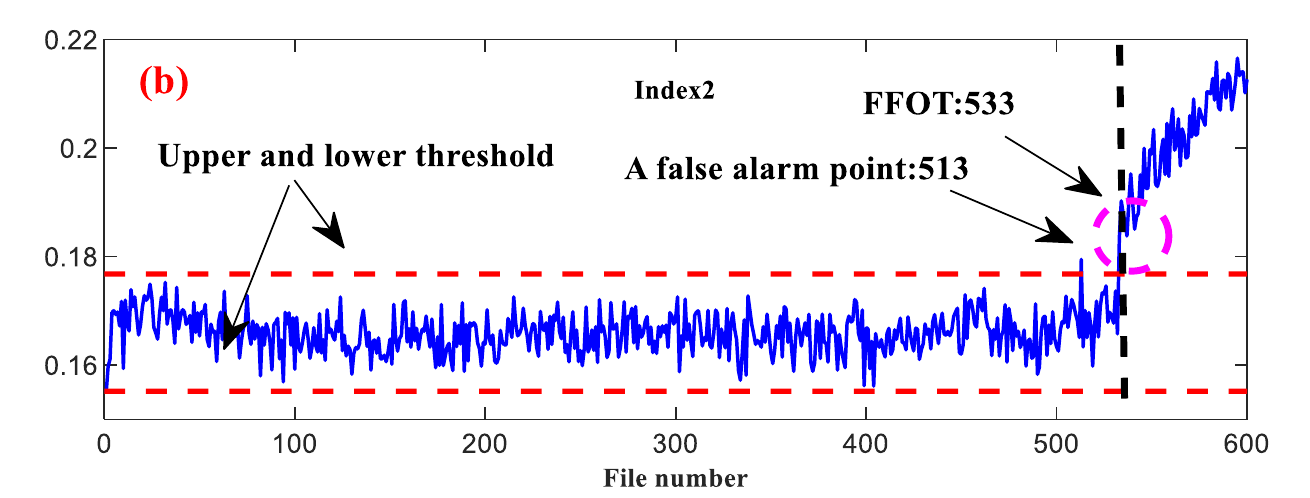}
		
	}%

	\subfigure{
		
		\includegraphics[width=15cm,height=5cm]{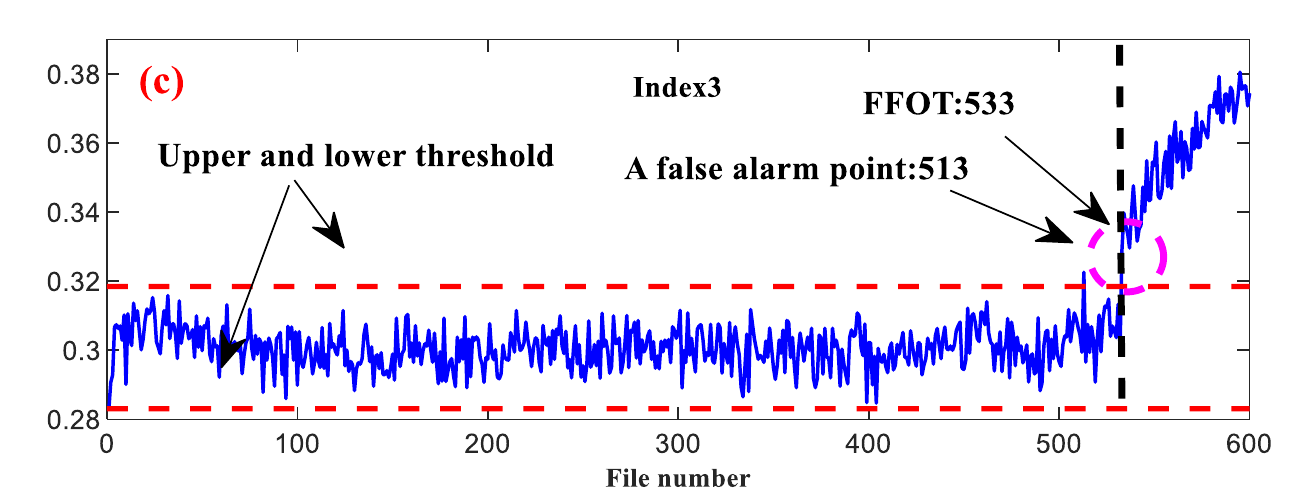}
		
	}%

	\subfigure{
		
		\includegraphics[width=15cm,height=5cm]{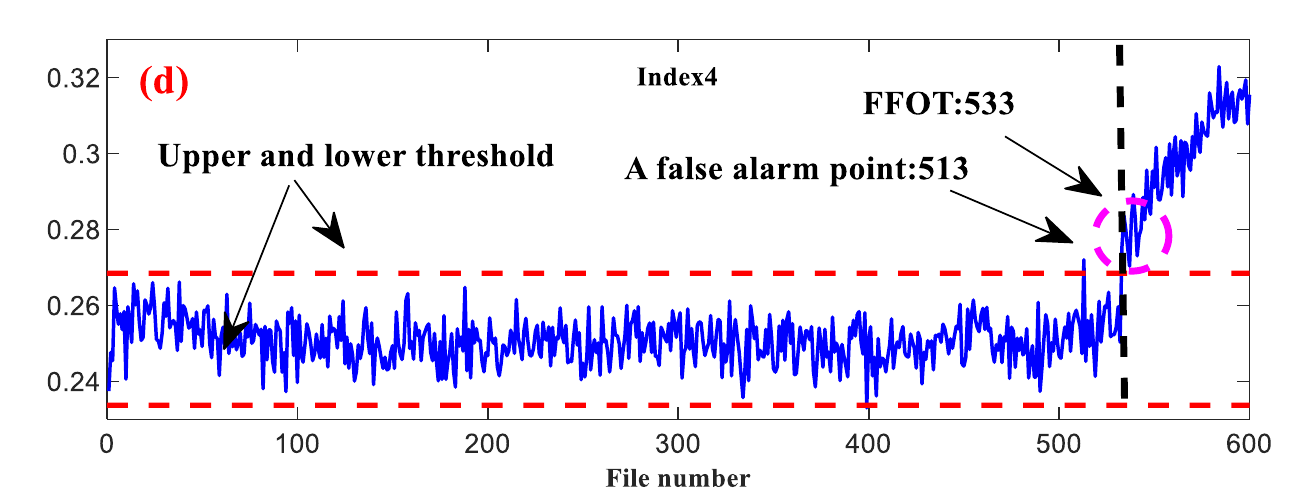}
		
	}%
	
	\centering
	\caption{First fault occurrence time determination and threshold analysis by three-sigma rule: (a) Index1; (b) Index2; (c) Index3; (d) Index4.}\label{FFOT}
\end{figure}

Hou et al. \cite{Hou2021b} analyzed the healthy state of the first to 300-th samples of the dataset and used the Lilliefors test to ensure that the historical health data followed the Gaussian distribution in their previous work. Therefore, $3\sigma$ rule can be used to establish statistical thresholds for the detection of the first failure occurrence time and to determine upper and lower thresholds for the division of health and fault states.
As marked by the red dotted lines and magenta dotted circles in Fig. \ref{FFOT}, our proposed improved indexes with their associate thresholds can effectively detecting FFOT at file No.533.
Based on the above analysis, some conclusions can be drawn. The designed HIs show a better monotonic degradation trend than the original measures with a more pronounced detection of FFOT and effectively suppress impulsive noise. This is mainly reflected in two aspects, firstly, the degradation curve is smoother, and secondly, the detection of fault-induced repetitive transients is more sensitive.

\subsubsection{Incipient Fault Diagnosis Around FFOT}\

\begin{figure}[!t]
	\centerline{\includegraphics[width=10.7cm,height=8cm]{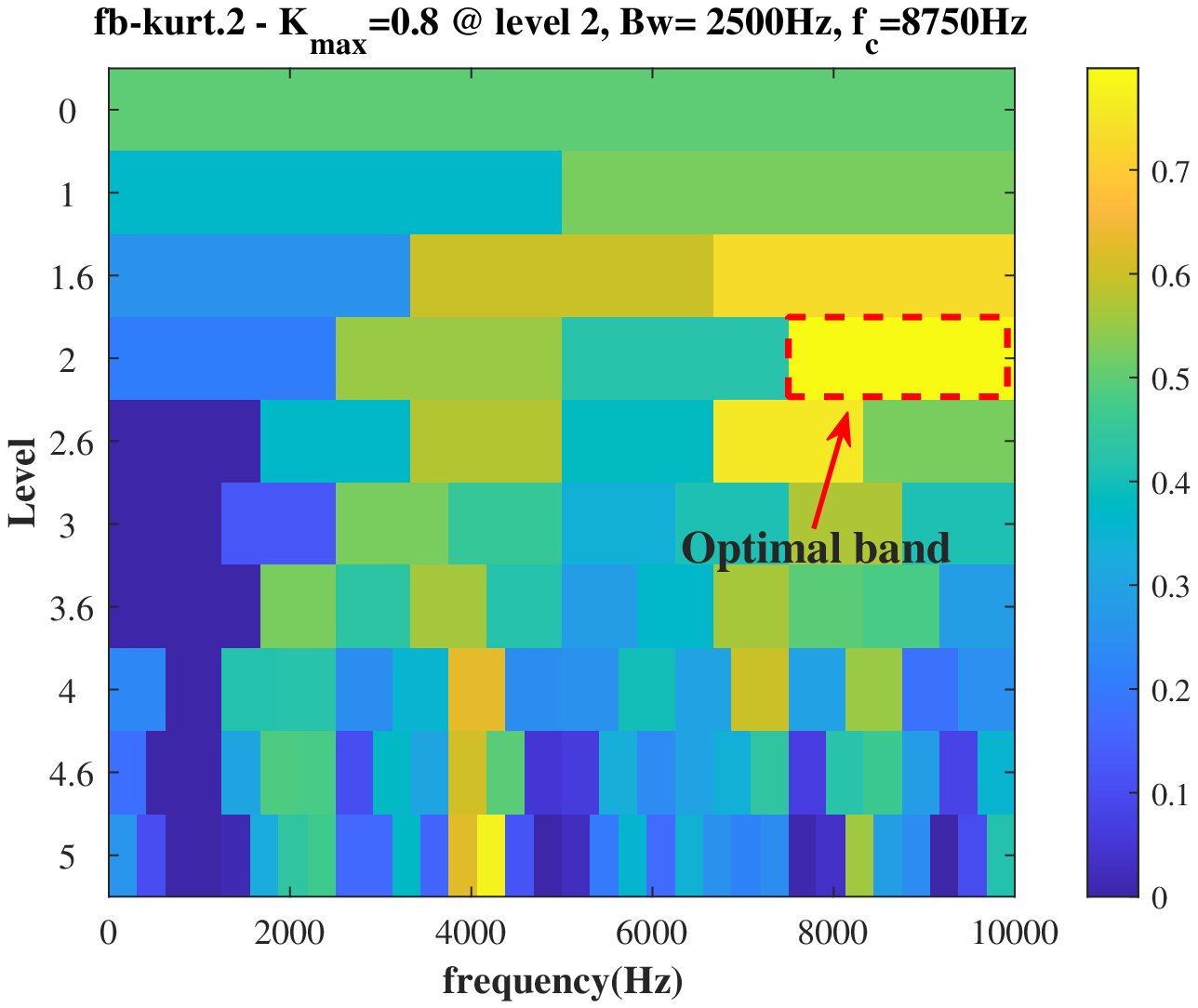}}
	\caption{The optimal filter band obtained by Fast Kurtogram.}
	\label{FK}
	
\end{figure}

\begin{figure}[!t]
	\centerline{\includegraphics[width=\columnwidth]{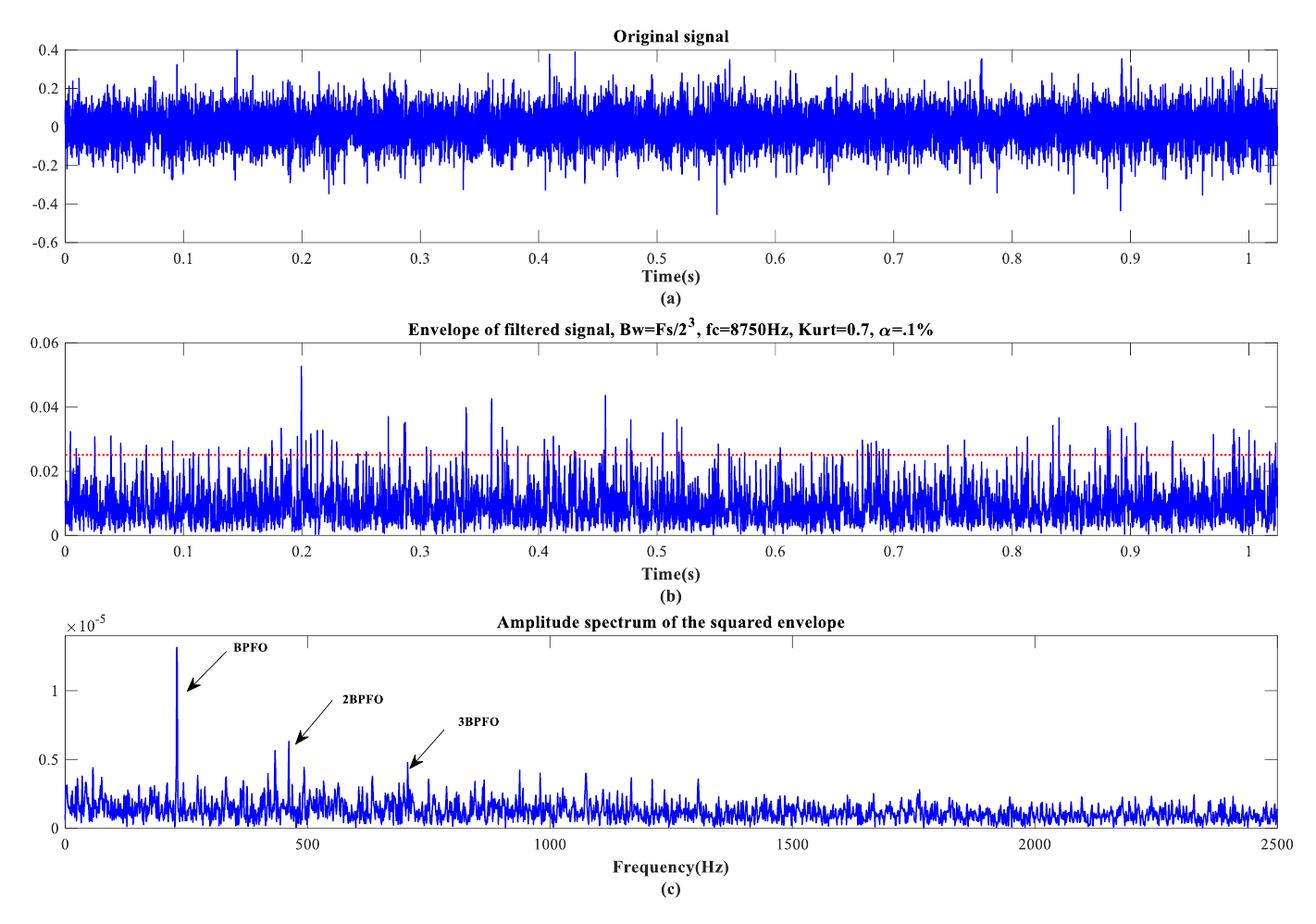}}
	\caption{The analysis results by envelope demodulation: (a)Original signal; (b)Envelope of filtered signal; (c)Amplitude spectrum of squared envelope.}
	\label{FKsignal}
\end{figure}

To verify correctness of indication FFOT based on designed HIs, the bearing fault signal at file No.534 was selected for incipient fault diagnosis. Fast Kurtogram method can help us to determine the optimal filter band center and band frequency, and then perform resonant envelope demodulation to extract the early fault characteristics of the bearing. The optimal filter band results are shown in Fig. \ref{FK}, and the envelope demodulation results are shown in Fig. \ref{FKsignal}. The final envelope demodulation results can clearly find the bearing outer ring fault frequency of 263.4Hz and its multiples, indicating that the bearing enters the fault state from the healthy state.

\subsubsection{Isolated-Forest Based Assessment of Degradation Stages}\

\begin{figure}[htbp]
	\centering
	\subfigure{
		
		\includegraphics[width=15cm,height=5cm]{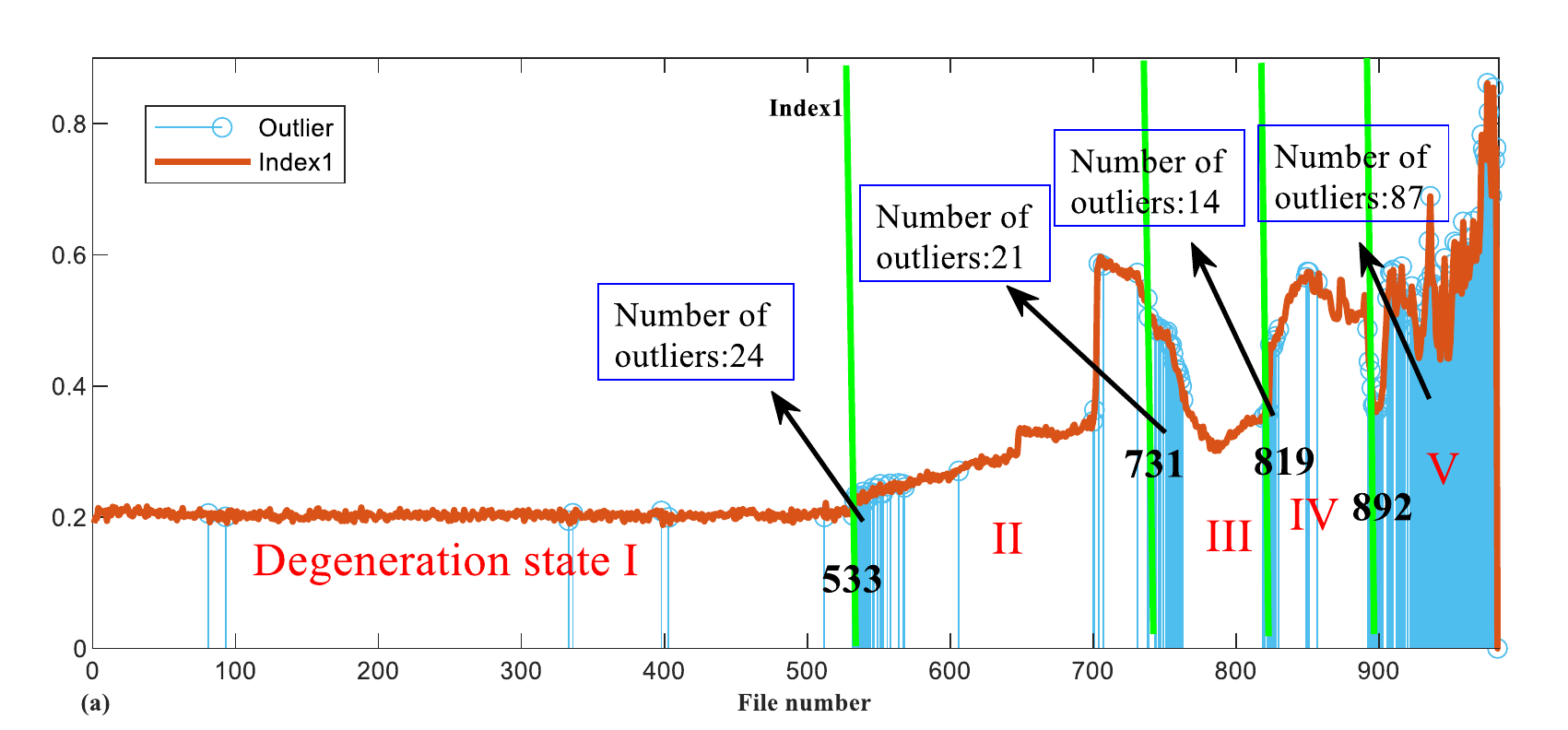}
	}%

	\subfigure{
		
		\includegraphics[width=15cm,height=5cm]{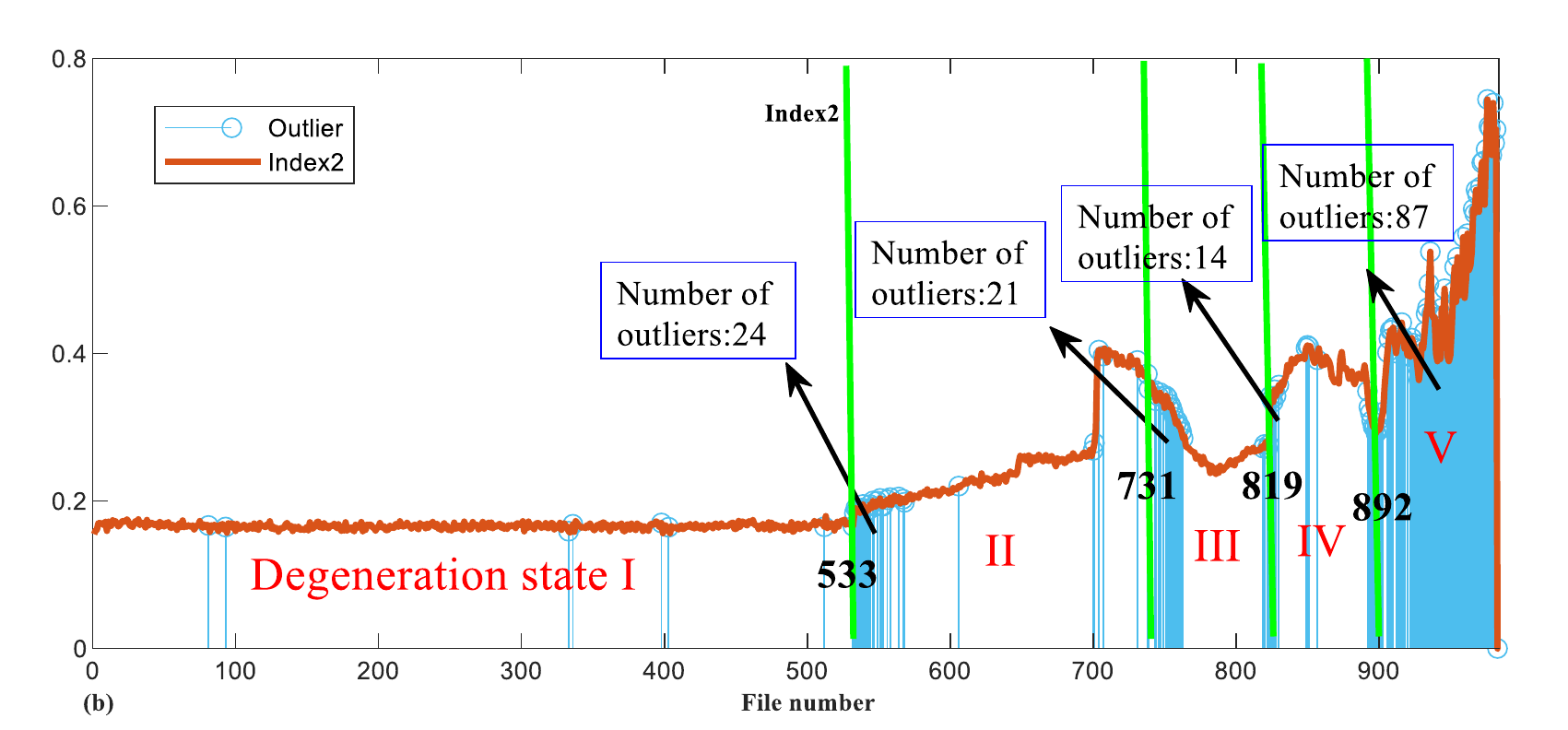}
		
	}%

	\subfigure{
		
		\includegraphics[width=15cm,height=5cm]{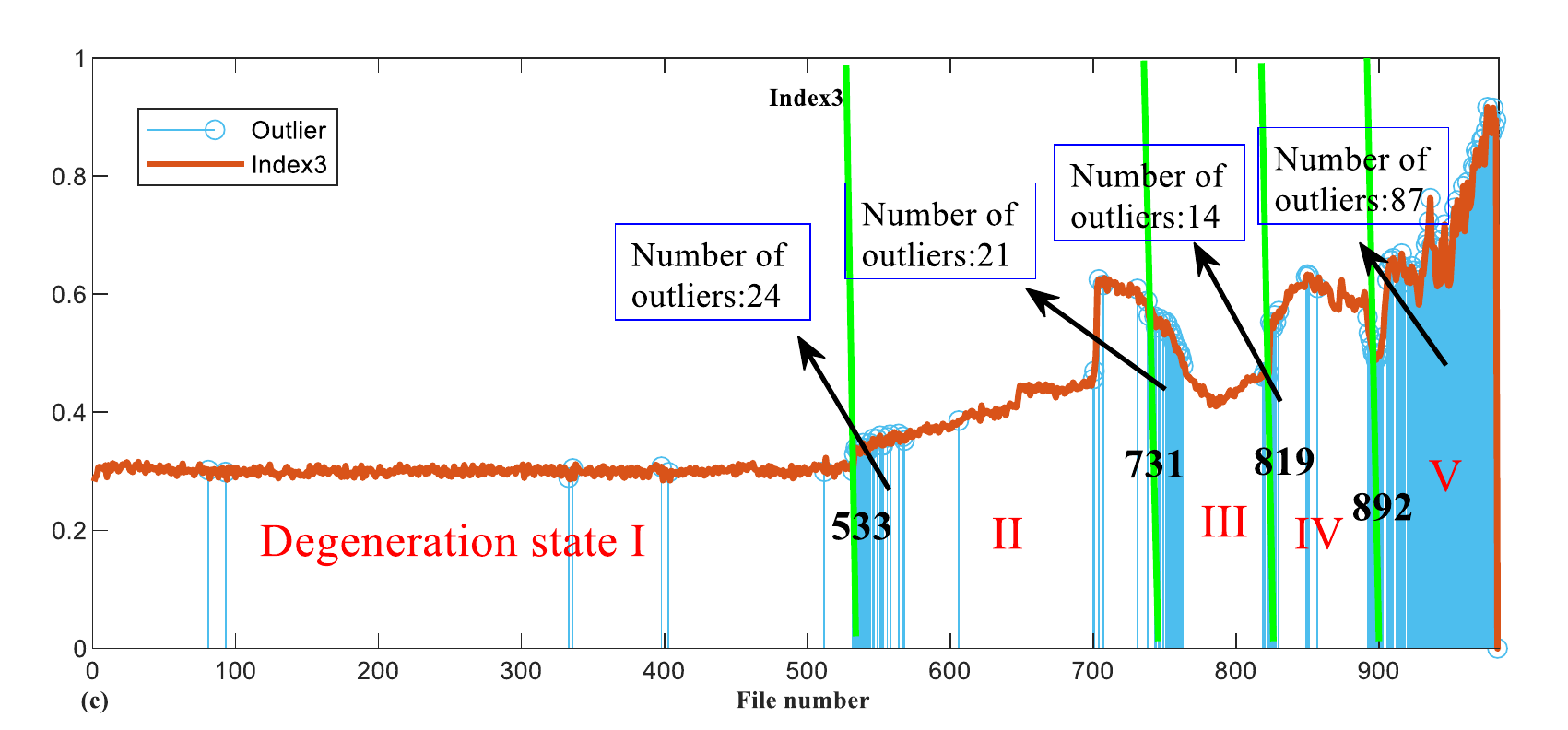}
		
	}%

	\subfigure{
		
		\includegraphics[width=15cm,height=5cm]{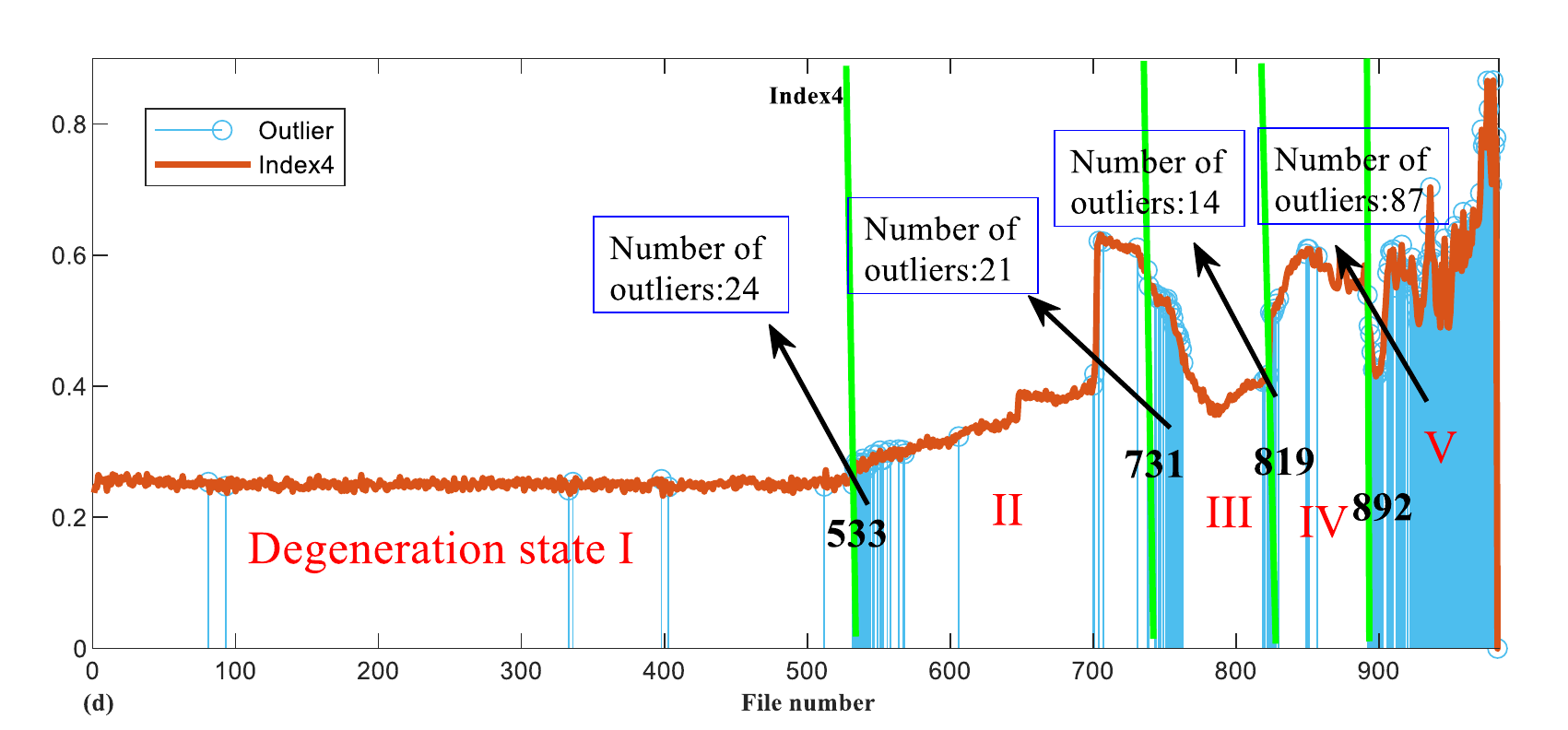}
		
	}%
	
	\centering
	\caption{Results of degradation stages assessment by Isolated-Forest algorithm applied on different HIs: (a) Index1; (b) Index2; (c) Index3; (d) Index4.}\label{IF}
\end{figure}

In this subsection, we combine the unsupervised isolated forest algorithm with newly designed HIs for the assessment of different degradation states of rolling bearings. Isolated forest algorithm was proposed by Liu et al. \cite{Liu2012} for anomaly detection. Isolated forest is suitable for anomaly detection of continuous numerical data, defining anomalies as "outliers that are more likely to be separated".
We consider that the degradation state of a machine changes when there are multiple consecutive outliers within a certain time interval. Specific hyperparameter settings of isolated forest algorithm are shown in Table \ref{IFtable}. The analysis results of four HIs are respectively shown in Fig. \ref{IF}.

\begin{table}[!h]
	\centering
	\caption{Specific Hyperparameter Settings of Isolated Forest Algorithm.}
	\label{IFtable}
	\begin{tabularx}{\columnwidth}{Xcccccc}
		\hline
		\hline
		\textbf{Hyperparameter}          & \textbf{Value}  \\
		\hline
		Maximum upper bound of outlier   & 10         \\
		The number of base estimators: n\_estimarors  & 256  \\
		The number of samples: max\_sample & 0.5 \\
		The number of features: max\_features & 1.0  \\
		Controls the pseudo-randomness: random\_state & 42 \\
		The number of jobs: n\_jobs & -1 \\
		Controls the verbosity: verbose & 0 \\
		\hline
		\hline
	\end{tabularx}
	\vspace{-0.25cm}
\end{table}

From the results of isolated forest analysis, it can be seen that five different degradation states were identified in the whole process of bearing degradation. The first state change occurs at file No.533 when the health state changes to the incipient fault state, which is consistent with the results analyzed in the previous subsections. A large number of continuous outliers began to appear after file No.892 so that it can be considered that the bearing entered a complete failure state starting from file No.892. This conclusion is basically consistent with that analyzed by RMS metric. Recognizing machine degradation state from the perspective of outliers is a new attempt in this paper. It is helpful to study the degradation mode of similar machines under the same operating condition and to further study the prediction of remaining useful life. However, how to further quantitatively evaluate the degree of machine degradation in different degradation states is a direction of future research, which will not be discussed in this paper.

\subsection{Case Study II: XJTU-SY Run-To-Failure Bearing Dataset}\

\begin{figure}[!t]
	\centerline{\includegraphics[width=9.5cm,height=6cm]{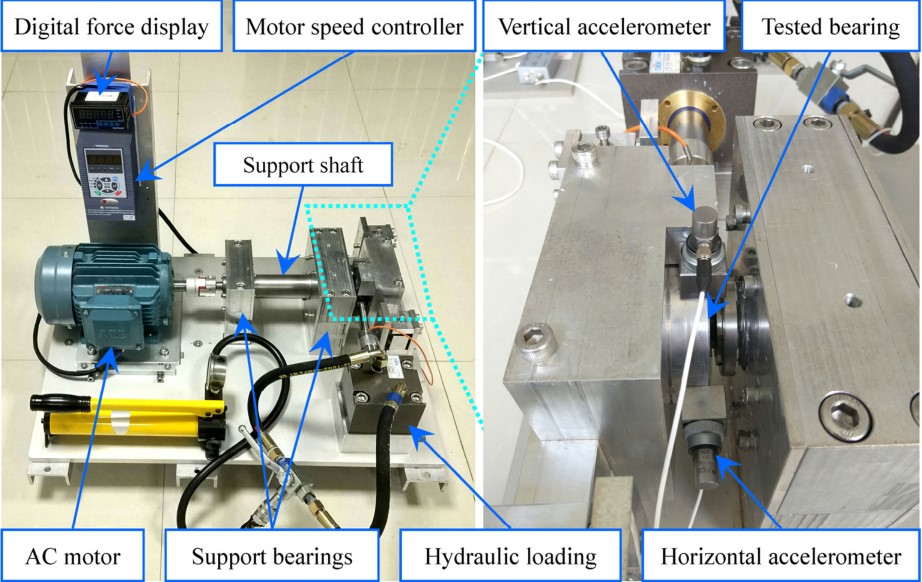}}
	\caption{The tested bearing platform of the XJTU-SY bearing dataset.}
	\label{xjd platform}
\end{figure}

To further validate the effectiveness and generalization of our proposed framework, this section is validated on the XJTU-SY run-to-failure bearing dataset. The bearing testbed is shown in Fig. \ref{xjd platform}. The sampling frequency is set at 25.6 kHz, and 32768 points (1.28 seconds) are collected each time with an interval of 1 minute. Details are described in \cite{Wang2020a}.

\begin{figure}[htbp]
	\centering
	\subfigure{
		\begin{minipage}[t]{0.5\linewidth}
			\centering
			\includegraphics[width=9cm,height=6.74cm]{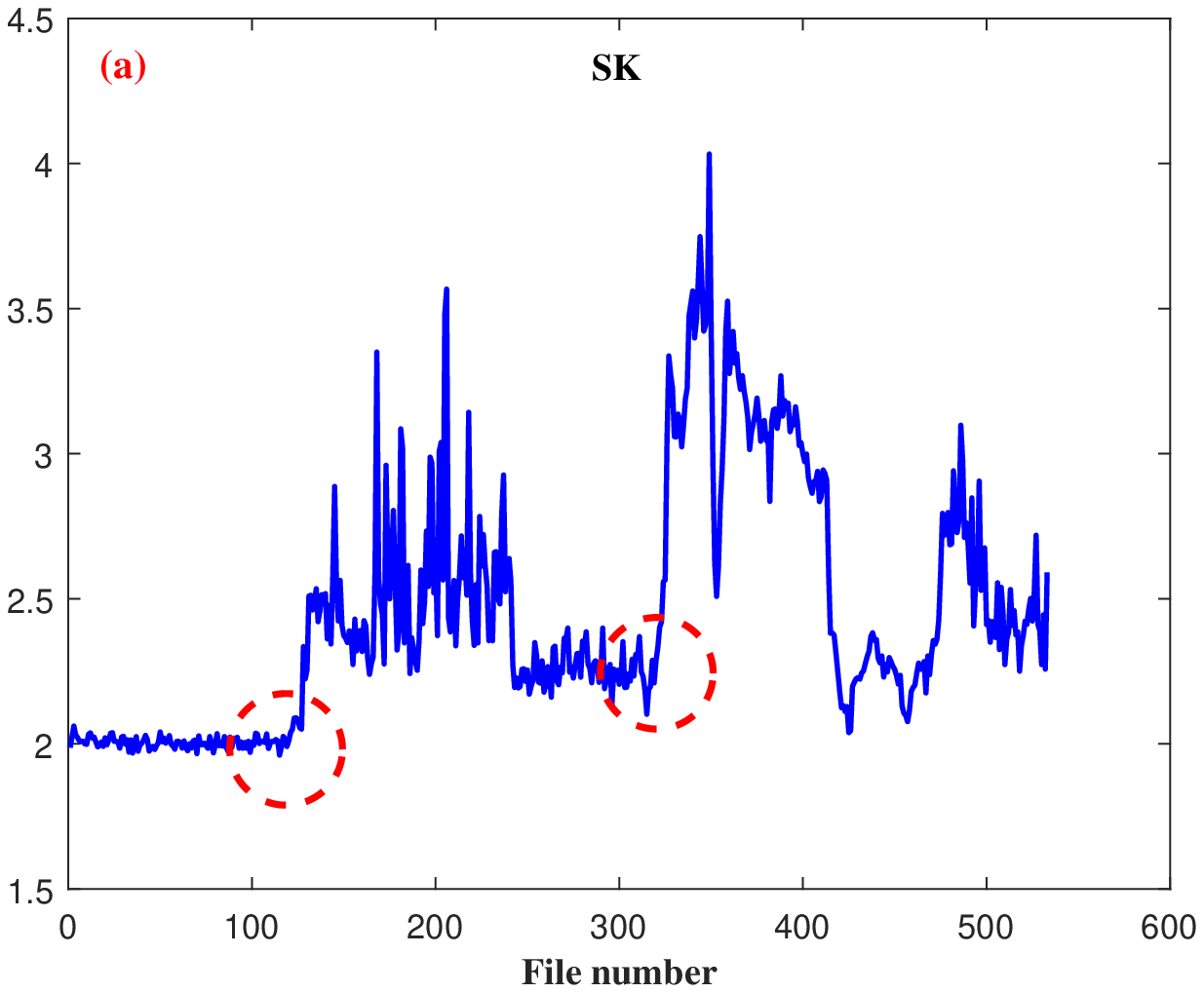}
		\end{minipage}%
	}%
	\subfigure{
		\begin{minipage}[t]{0.5\linewidth}
			\centering
			\includegraphics[width=9cm,height=6.74cm]{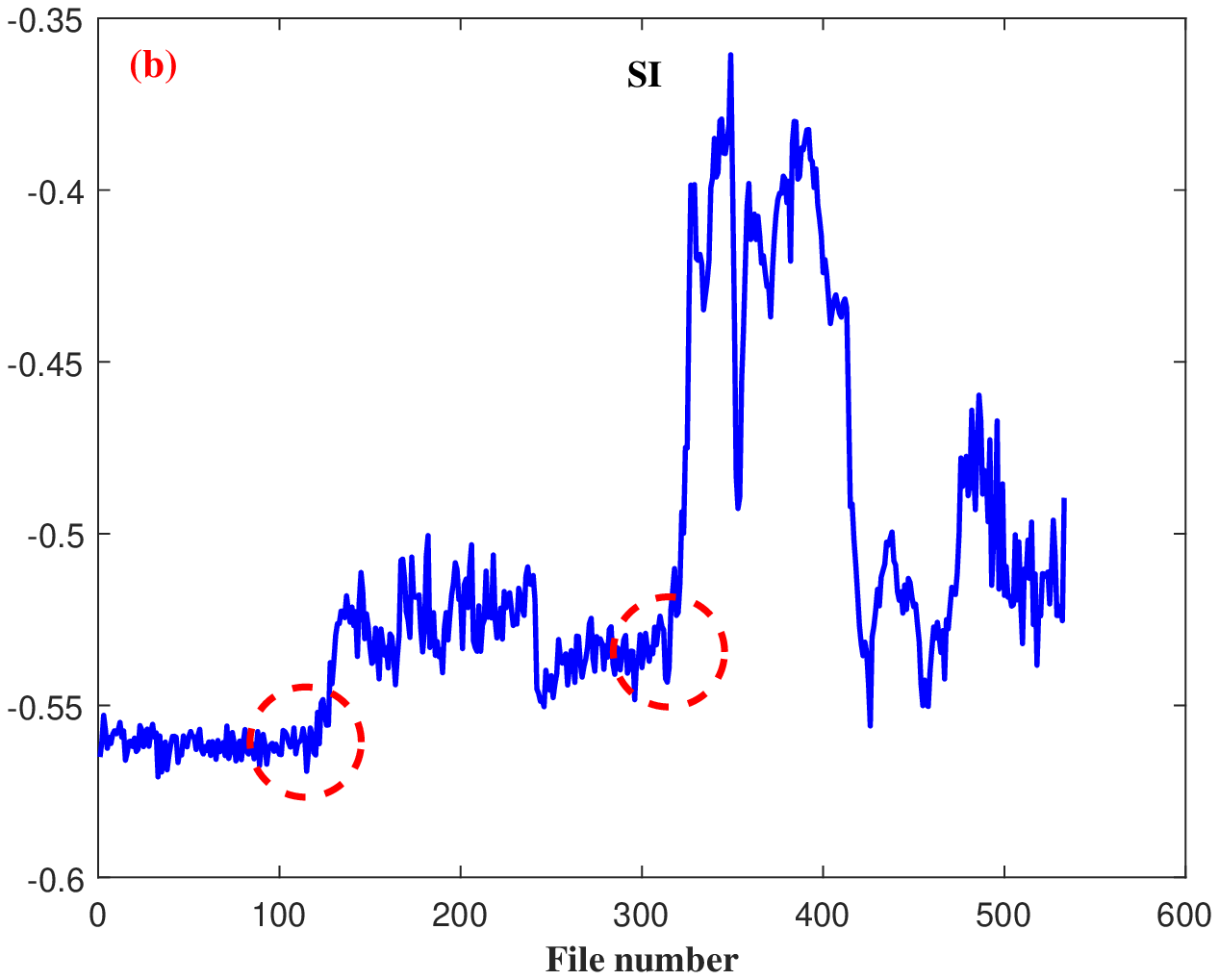}
		\end{minipage}%
	}%

	\subfigure{
		\begin{minipage}[t]{0.5\linewidth}
			\centering
			\includegraphics[width=9cm,height=6.74cm]{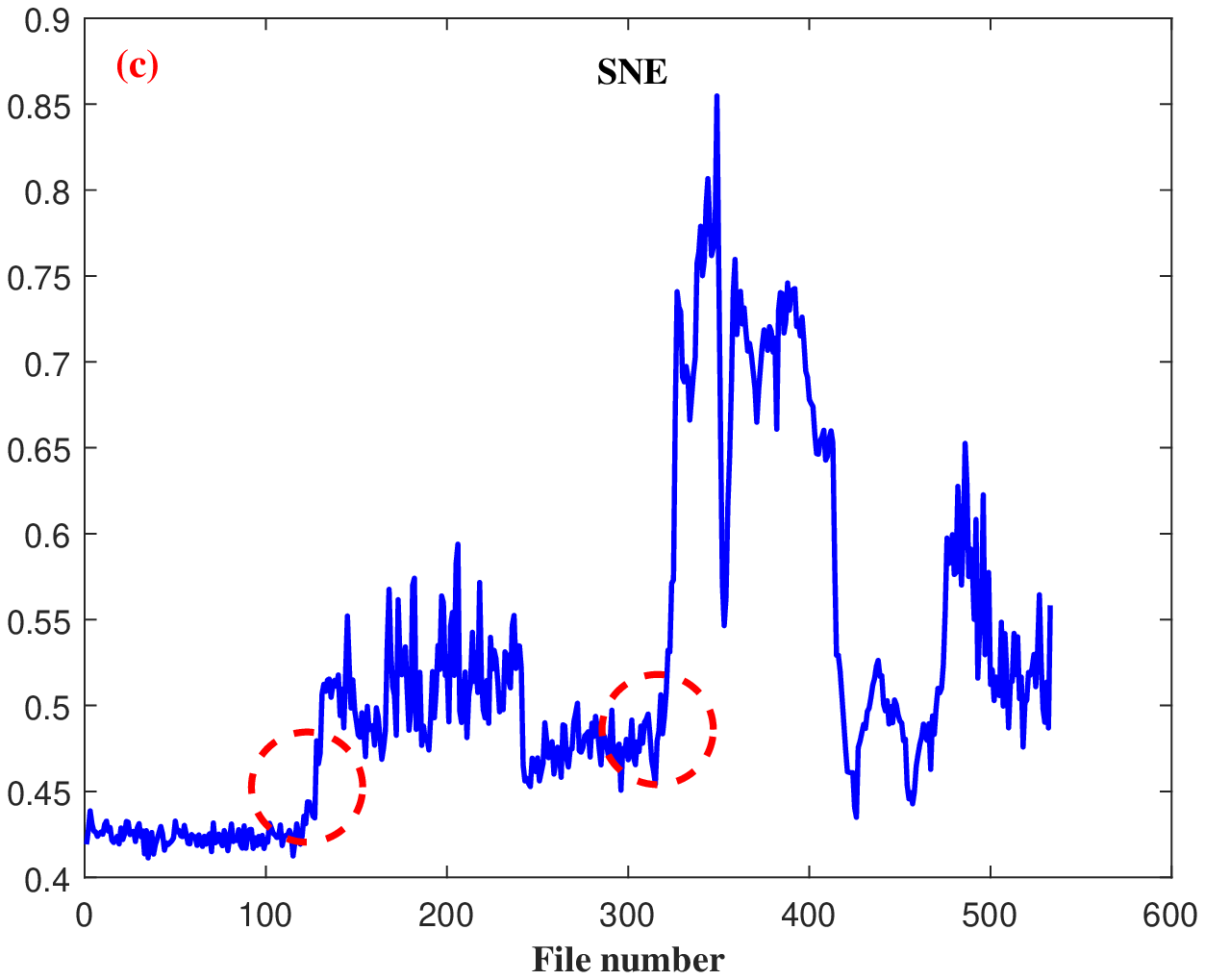}
		\end{minipage}%
	}%
	\subfigure{
		\begin{minipage}[t]{0.5\linewidth}
			\centering
			\includegraphics[width=9cm,height=6.74cm]{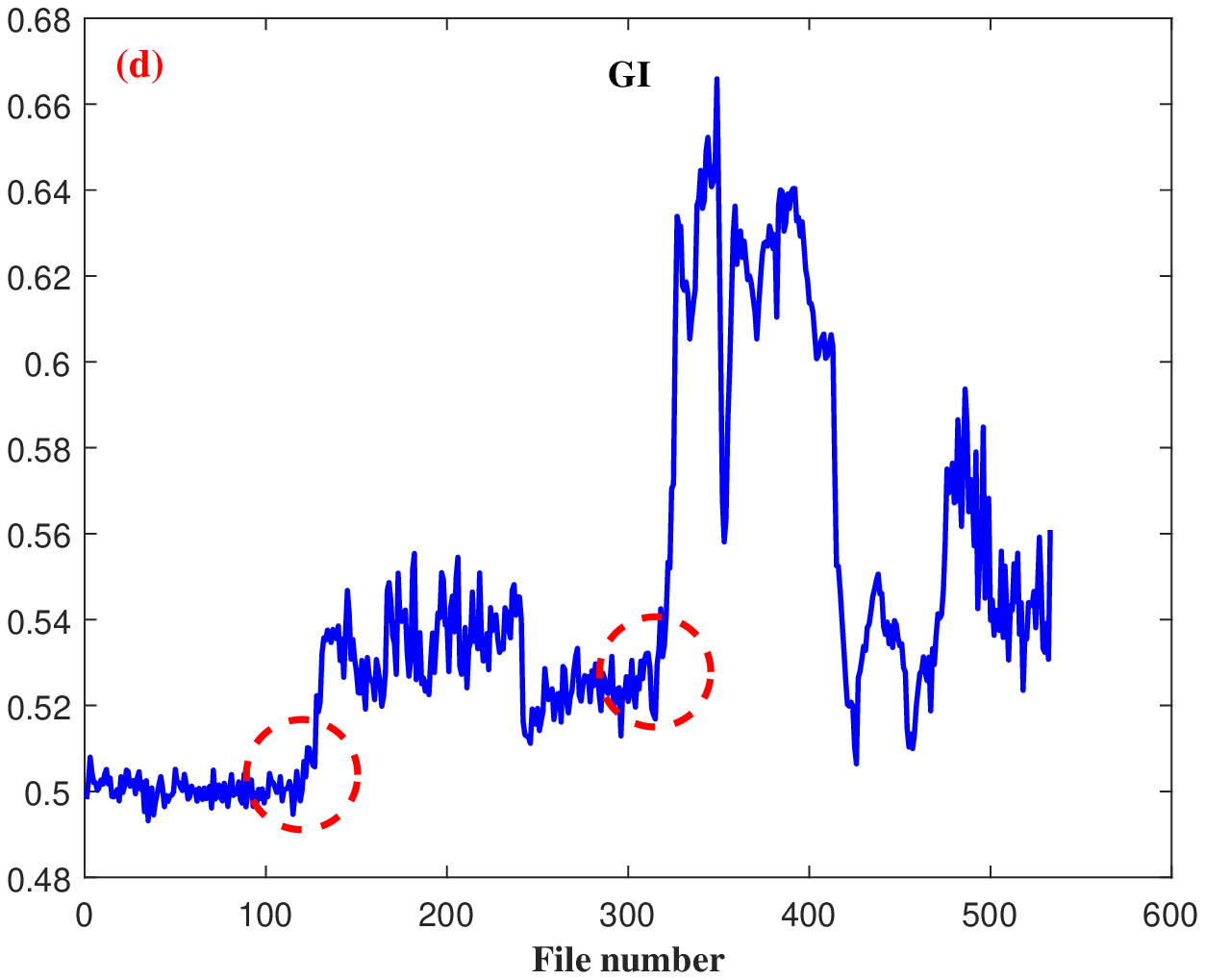}
		\end{minipage}%
	}%
	
	\centering
	\caption{Bearing degradation curves plotted by classic sparsity measures: (a) spectral kurtosis; (b) spectral smoothness index; (c) spectral negative entropy; (d) Gini index.}\label{xjdclassic}
\end{figure}

\begin{figure}[htbp]
	\centering
	\subfigure{
		\begin{minipage}[t]{0.5\linewidth}
			\centering
			\includegraphics[width=9cm,height=6.74cm]{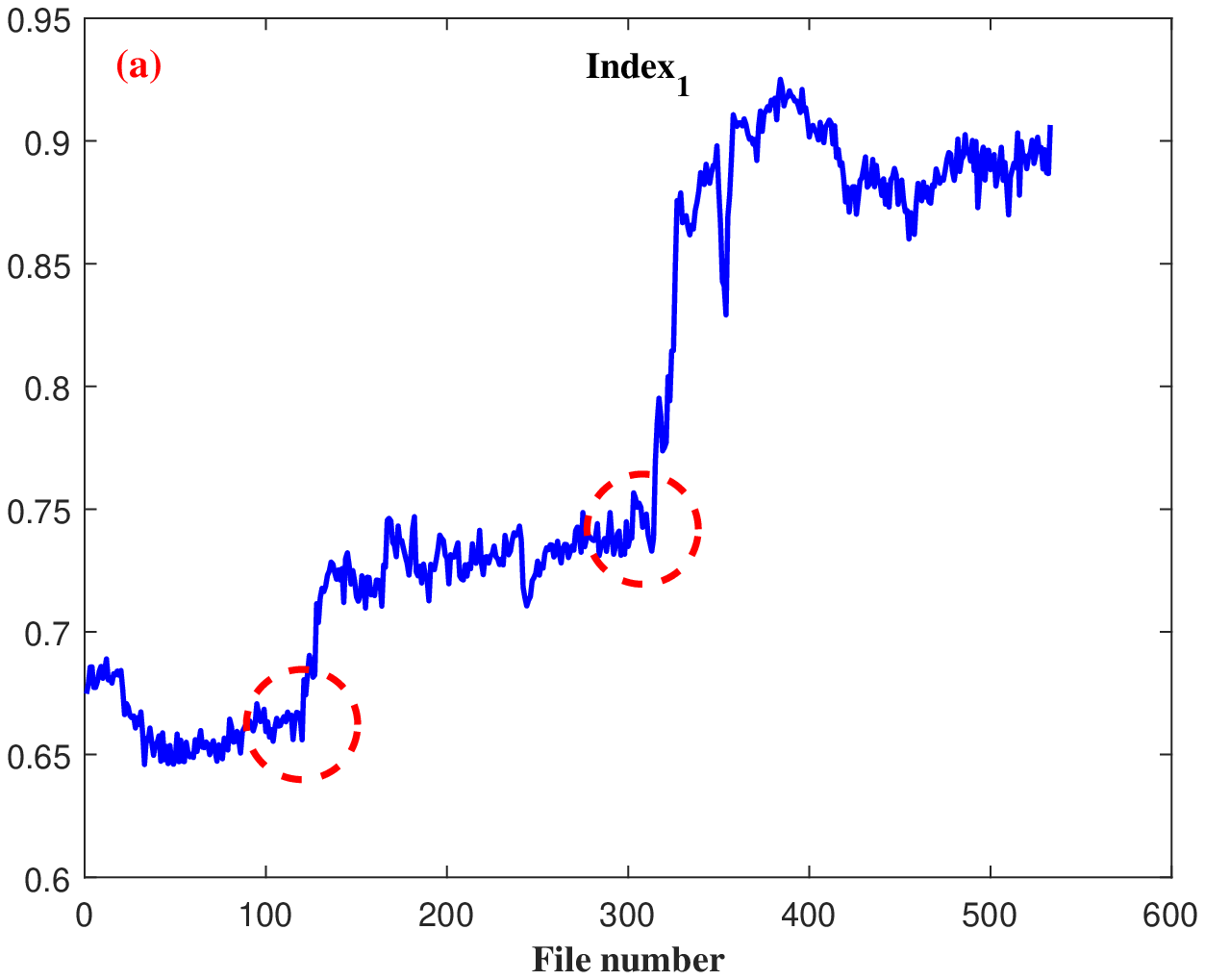}
		\end{minipage}%
	}%
	\subfigure{
		\begin{minipage}[t]{0.5\linewidth}
			\centering
			\includegraphics[width=9cm,height=6.74cm]{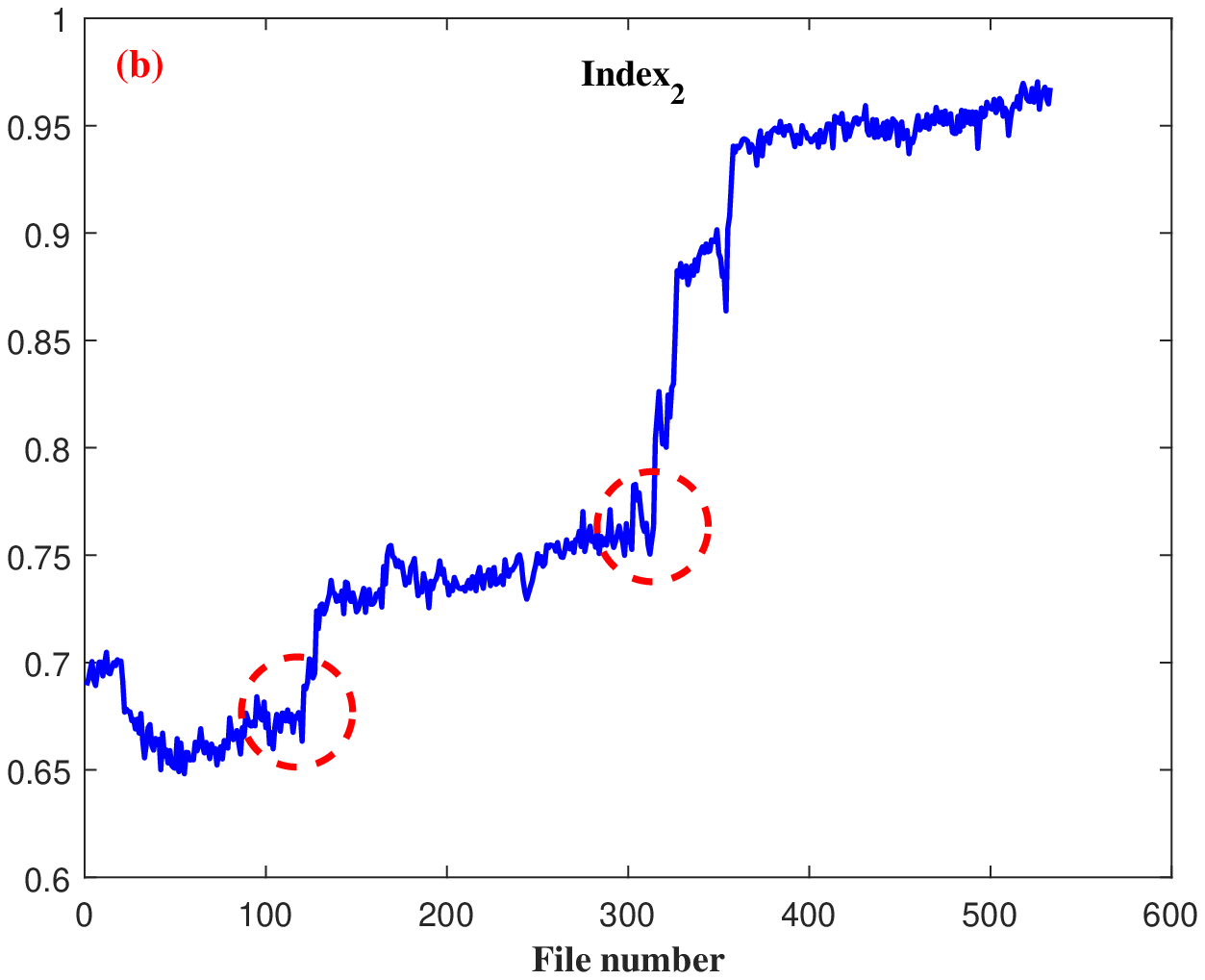}
		\end{minipage}%
	}%

	\subfigure{
		\begin{minipage}[t]{0.5\linewidth}
			\centering
			\includegraphics[width=9cm,height=6.74cm]{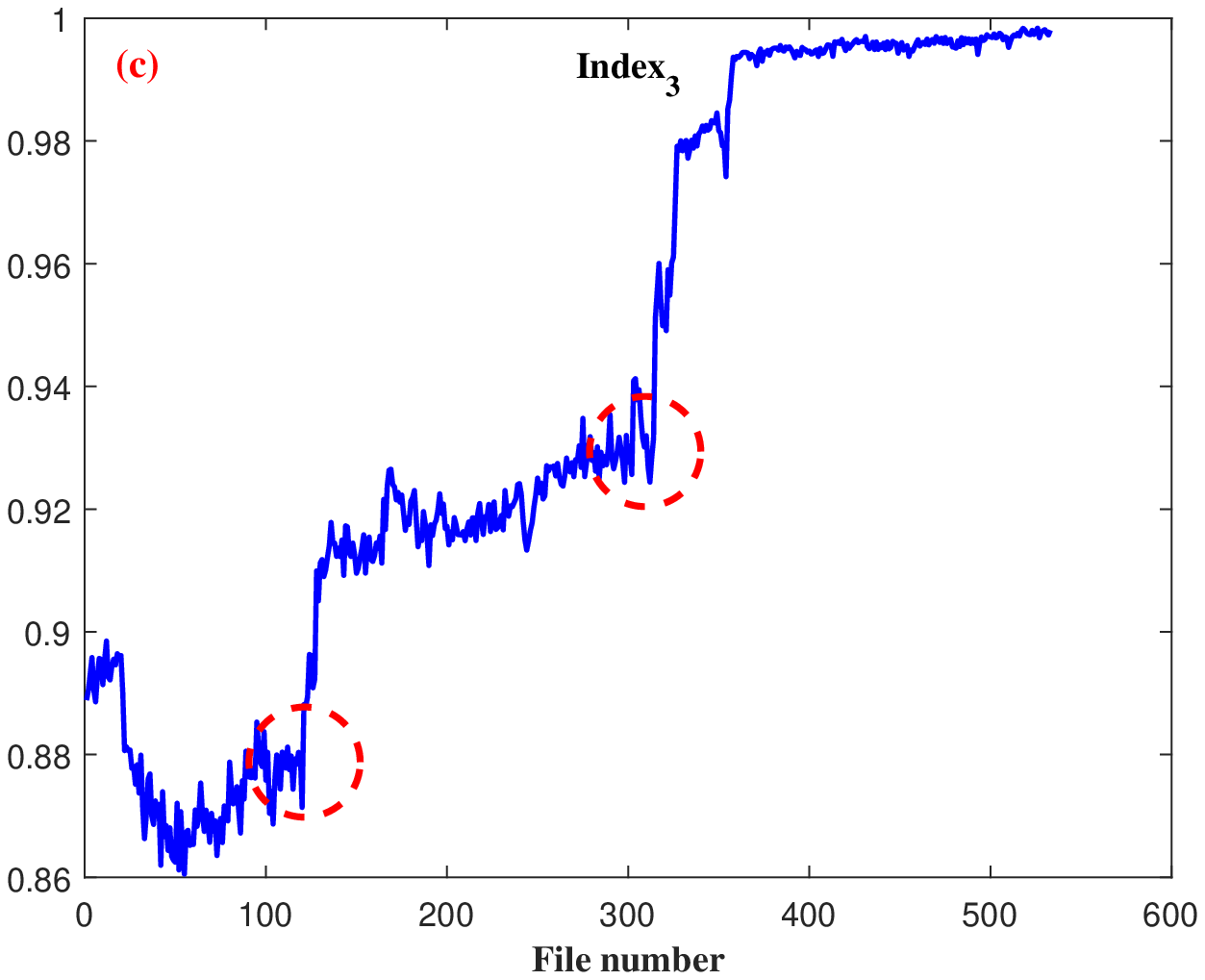}
		\end{minipage}%
	}%
	\subfigure{
		\begin{minipage}[t]{0.5\linewidth}
			\centering
			\includegraphics[width=9cm,height=6.74cm]{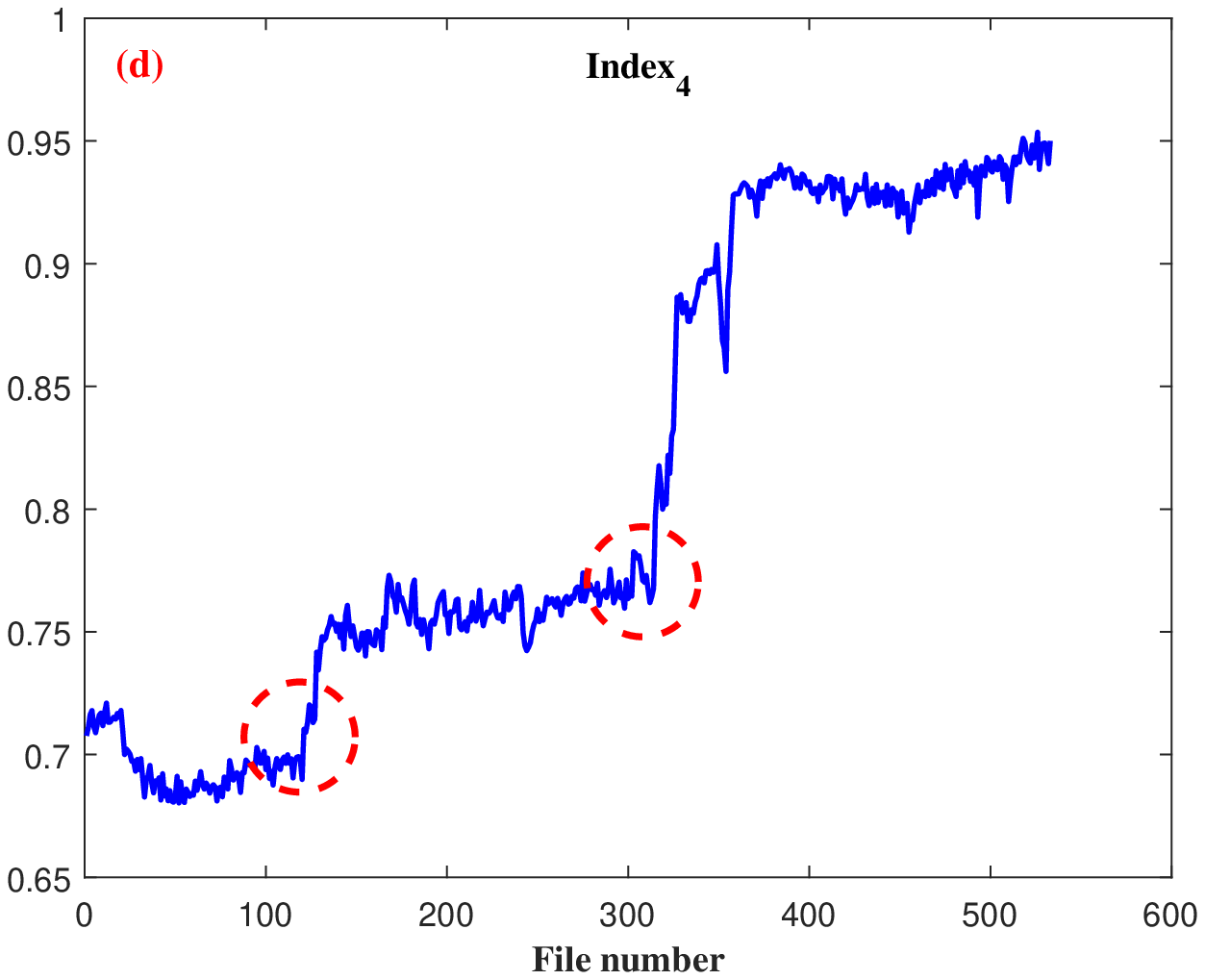}
		\end{minipage}%
	}%
	
	\centering
	\caption{Bearing degradation curves plotted by: (a) Index1 ; (b) Index2; (c) Index3; (d) Index4.}\label{xjdindex}
\end{figure}

The curves of bearing degradation quantified by classic spectral kurtosis, spectral smoothness index, spectral negative entropy and Gini index are plotted in Fig. \ref{xjdclassic}. Meanwhile, the curves of bearing degradation quantified by $HI_1$-$HI_4$ are plotted in Fig \ref{xjdindex}. From the above figures, it can be concluded that the degradation trends quantified by four classic sparsity measures are poor and can only roughly identify the moment of incipient failure of bearings. The degradation curves quantified by $HI_1$-$HI_4$ can show good degradation trends and own the ability to identify incipient and severe bearing faults. In terms of fluctuation, except for degradation curves quantified by $HI_3$, the vast majority of HIs show less fluctuant. The validity of the proposed indexes and generalization of the framework based on ratios of multivariate power mean functions are further verified by the XJTU-SY run-to-failure bearing dataset. Based on the general paradigm, more reasonable and advanced health indexes can be easily designed, and the specific index design formulas can be determined depending on the specific conditions of industrial sites.

\section{Conclusion}
In this paper, we have studied the sparse attributes of smoothness index and negative entropy and given detailed proofs. To develop a general method for construction of health indexes with monotonicity that enables incipient fault detection, multivariate power mean functions (MPMFs) were introduced and investigated. Then, it is found that the existing classical sparsity measures can be reformulated as ratios of different MPMFs, including spectral kurtosis, spectral smoothness index, spectral negative entropy and Gini index. Finally, the general health index construction paradigm based on ratios of MPMFs are proposed for family expansion of sparsity measures. Four specific health indexes are designed as examples, and their sparse attributes and other properties are analyzed in detail. Their ability to monitor health of machines has been verified through two run-to-life bearings datasets.

\appendix

\section{}

\subsection{Proofs Of Smoothness Index}

\subsubsection{SI and Nonnegativity}\ 

\textit{Theorem A.1}: SI is non-negative.
\begin{equation}
	SI(\overset\rightarrow{x})\geq0
\end{equation}

\textit{Proof}: As each element in vector $\mathbf{x}=[x_1,x_2,\dots,x_N]$ is non-negative. Clearly, SI is non-negative.

\subsubsection{SI and D1}\

\textit{Theorem A.2}: SI does not satisfy D1.
\begin{equation*}
	SI \left ( \left [x_1 \dots x_i-\alpha \dots x_j+\alpha \dots \right ] \right )<SI(\overset\rightarrow{x})
\end{equation*}
for all $\alpha$, $x_i$, $x_j$ such that $x_i>x_j$ and $0<\alpha<\frac{x_i-x_j}{2}$.

\textit{Proof}: As $S I=\frac{\sqrt[N]{\prod_{i=1}^{N} x_{i}}}{\sum_{i=1}^{N} x_{i} / N}$we can restate the above as
\begin{equation*}
	\frac{\sqrt{\prod_{k \neq i, j}^{N} x_{k}\left(x_{i}-\alpha\right)\left(x_{j}+\alpha\right)}}{\left(\left(\sum_{k=1}^{N} x_{k}\right)+\alpha-\alpha\right) / N}<\frac{\sqrt{\prod_{i=1}^{N} x_{k}}}{\sum_{k=1}^{N} x_{k} / N} 
\end{equation*}

This simplifies to
\begin{equation*}
	\sqrt{\prod_{k \neq i, j}^{N} x_{k}\left(x_{i}-\alpha\right)\left(x_{j}+\alpha\right)}<\sqrt{\prod_{i=1}^{N} x_{k}}
\end{equation*}
\begin{equation*}
	x_{1} x_{2} \cdots\left(x_{i}-\alpha\right) \cdots\left(x_{j}+\alpha\right) \cdots x_{N}<x_{1} x_{2} \cdots x_{N}
\end{equation*}
\begin{equation*}
	\left(x_{i}-\alpha\right)\left(x_{j}+\alpha\right)<x_{i} x_{j}
\end{equation*}
\begin{equation*}
	x_{i}-x_{j}-\alpha<0
\end{equation*}
which we konw is not true as $x_i>x_j$ and $0<\alpha<\frac{x_i-x_j}{2}$.

\subsubsection{SI and D2}\

\textit{Theorem A.3}: SI satisfies D2.
\begin{equation*}
	SI(\alpha \overset\rightarrow{x})=SI(\overset\rightarrow{x}),\forall \alpha \in \mathbb{R}, \alpha>0
\end{equation*}

\textit{Proof}:
\begin{equation*}
	SI(\alpha \vec{x})=\frac{\sqrt[N]{\prod_{i=1}^{N} \alpha x_{i}}}{\sum_{i=1}^{N} \alpha x_{i} / N}=\frac{\alpha \sqrt{\prod_{i=1}^{N} x_{i}}}{\alpha \sum_{i=1}^{N} x_{i} / N}=SI(\vec{x})
\end{equation*}

\subsubsection{SI and D3}\

\textit{Theorem A.4}: SI does not satisfy D3.
\begin{equation*}
	SI(\alpha+\overset\rightarrow{x})<SI(\overset\rightarrow{x}),\forall \alpha \in \mathbb{R}, \alpha>0
\end{equation*}

\textit{Proof}: Set\begin{equation*}
	F(\alpha)=\frac{\sqrt[N]{\prod_{i=1}^{N}\left(x_{i}+\alpha\right)}}{\left(\sum_{i=1}^{N} x_{i} / N\right)+\alpha}=\frac{\sqrt[N]{A(\alpha)}}{B(\alpha)}
\end{equation*}
\begin{equation*}
	\ln [A(\alpha)]=\ln \left[\left(\alpha+x_{1}\right)\left(\alpha+x_{2}\right) \cdots\left(\alpha+x_{N}\right)\right]
\end{equation*}
Derivation on both sides of the equation
\begin{equation*}
	\frac{A^{\prime}(\alpha)}{A(\alpha)}=\sum_{i=1}^{N} \frac{1}{\alpha+x_{i}}
\end{equation*}
\begin{equation*}
	A^{\prime}(\alpha)=A(\alpha) \sum_{i=1}^{N} \frac{1}{\alpha+x_{i}}
\end{equation*}
It follows that
\begin{equation*}
	\frac{\partial F}{\partial \alpha}=\frac{\frac{1}{N} A(\alpha)^{\frac{1}{N}^{-1}} A^{\prime}(\alpha)}{B^{2}(\alpha)}=\frac{\frac{1}{N} A(\alpha)^{\frac{1}{N}} \sum_{i=1}^{N} \frac{1}{\alpha+x_{i}}-A(\alpha)^{\frac{1}{N}}}{B^{2}(\alpha)}
\end{equation*}
We can ignore the denominator as it is clearly positive.It is clear that $\frac{\partial F}{\partial \alpha}$ could be positive. Hence SI does not satisfy D3.

\subsubsection{SI and D4}\

\textit{Theorem A.5}: SI satisfies D4.
\begin{equation*}
	SI(\vec{x})=SI(\vec{x} \| \vec{x})=SI(\vec{x}\|\vec{x}\| \vec{x})=SI(\vec{x}\|\vec{x}\| \cdots \| \vec{x})
\end{equation*}

\textit{Proof}:
\begin{equation*}
	\operatorname{SI}(\underbrace{\vec{x}\|\vec{x} \cdots\| \vec{x}}_{m})=\frac{\sqrt[m N]{\prod_{i=1}^{N} x_{i}^{m}}}{\left(m \sum_{i=1}^{N} x_{i}\right) / m N}=\frac{\sqrt[N]{\prod_{i=1}^{N} x_{i}}}{\sum_{i=1}^{N} x_{i} / N}=\operatorname{SI}(\vec{x})
\end{equation*}

\subsubsection{SI and P1}\

\textit{Theorem A.6}: SI does not satisfy P1.

$\forall i$, $\exists \beta=\beta_{i}>0$, such that $\forall \alpha>0$
\begin{equation*}
	SI \left ( \left [x_1 \dots x_i+\beta+\alpha \dots  \right ] \right )>SI\left ( \left [x_1 \dots x_i+\beta \dots \right ] \right )
\end{equation*}

\textit{Proof}: Set
\begin{equation*}
	F(\beta)=\frac{\sqrt[N]{x_{i}+\beta}}{\sum_{i=1}^{N} x_{i}+\beta}
\end{equation*}
It follows that
\begin{equation*}
	\begin{aligned}
		\frac{\partial F}{\partial \beta}&=\frac{\left[\frac{1}{N}\left(x_{i}+\beta\right)^{\frac{1}{N}-1}\right]\left(\sum_{i=1}^{N} x_{i}+\beta\right)-\left(x_{i}+\beta\right)^{\frac{1}{N}}}{\left(\sum_{i=1}^{N} x_{i}+\beta\right)^{2}} \\& =\frac{\left[\frac{1}{N}\left(x_{i}+\beta\right)^{\frac{1}{N}-1}\right]\left[\sum_{i=1}^{N} x_{i}+\beta-N\left(x_{i}+\beta\right)\right]}{\left(\sum_{i=1}^{N} x_{i}+\beta\right)^{2}} \\
		=& \frac{\left[\frac{1}{N}\left(x_{i}+\beta\right)^{\frac{1}{N}-1}\right]}{\left(\sum_{i=1}^{N} x_{i}+\beta\right)^{2}}\left[(1-N) \beta+\sum_{i=1}^{N} x_{i}-N x_{i}\right]
	\end{aligned}
\end{equation*}
Clearly, for $\forall i$, $\frac{\partial F}{\partial \beta}$ could be negative. So SI does not satisfy P1.
\subsubsection{SI and P2}\

\textit{Theorem A.7}: SI does not satisfy P2.
\begin{equation*}
	SI(\vec{x} \| 0)>SI(\vec{x})
\end{equation*}

\textit{Proof}:
\begin{equation*}
	SI(\vec{x} \| 0)=0 \leq SI(\vec{x})
\end{equation*}

\subsection{Proofs Of Spectral Negative Entropy}
\subsubsection{SNE and Nonnegativity}\ 

\textit{Theorem B.1}: SNE is non-negative.
\begin{equation*}
	SNE(\overset\rightarrow{x})\geq0
\end{equation*}

\textit{Proof}: According to Gibbs' inequality, we can know that if $\sum_{i=1}^{n} p_{i}=\sum_{i=1}^{n} q_{i}=1, p_{i}, q_{i} \in(0,1]$, then $-\sum_{i=1}^{n} p_{i} \log p_{i} \leq-\sum_{i=1}^{n} p_{i} \log q_{i}$. Suppose $\bar{x}=\sum_{i=1}^{N} x_{i} / N$, $p_{i}=\frac{x_{i}}{N \bar{x}}$, $q_i=\frac{1}{N}$. It follows that
\begin{equation*}
	\begin{aligned}
		S N E(\vec{x})&=\frac{1}{N} \sum_{i=1}^{N} \frac{x_{i}}{\bar{x}} \ln \frac{x_{i}}{\bar{x}}=\sum_{i=1}^{N} \frac{x_{i}}{N \bar{x}} \ln \frac{x_{i}}{N \bar{x}}+\ln N \\&\geq \ln \frac{1}{N} \sum_{i=1}^{N} \frac{x_{i}}{N \bar{x}}+\ln N=0
	\end{aligned}
\end{equation*}Clearly, SNE is non-negative.

\subsubsection{SNE and D1}\ 

\textit{Theorem B.2}: SNE satisfies D1.
\begin{equation*}
	SNE \left ( \left [x_1 \dots x_i-\alpha \dots x_j+\alpha \dots \right ] \right )<SNE(\overset\rightarrow{x})
\end{equation*}
for all $\alpha$, $x_i$, $x_j$ such that $x_i>x_j$ and $0<\alpha<\frac{x_i-x_j}{2}$.

\textit{Proof}:
\begin{equation*}
	SNE\left(\left[x_{1} \cdots x_{i}-\alpha \cdots x_{j}+\alpha \cdots\right]\right)-SNE(\vec{x}) 
\end{equation*}
\begin{equation*}
	=\frac{x_{i}-\alpha}{A} \ln \frac{x_{i}-\alpha}{A}+\frac{x_{j}+\alpha}{A} \ln \frac{x_{j}+\alpha}{A}-\frac{x_{i}}{A} \ln \frac{x_{i}}{A}-\frac{x_{j}}{A} \ln \frac{x_{j}}{A}
\end{equation*}
Converted to Function $F(x)=xlnx$, in any interval $[x_{j}, x_{i}]$, the sum of the function values at the endpoints of the interval must be greater than the sum of the function values of two points in any interval $[x_{j}+a, x_{i}-a]$, which is symmetric about the midpoint of interval $[x_{j}, x_{i}]$. Clearly, $F^{\prime \prime}(x)>0$, and $F(x)$ is a lower convex function.
It follows that in any interval $[x_{j}, x_{i}]$
\begin{equation*}
	F\left(x_{i}\right)+F\left(x_{j}\right)>F\left(x_{i}-\alpha\right)+F\left(x_{j}+\alpha\right)
\end{equation*}
where $x_i>x_j$ and $0<\alpha<\frac{x_i-x_j}{2}$.
Define an arbitrary interval $[x_{j}, x_{i}]$ of length $2d$, $d>0$.

Set
\begin{equation*}
	H(d)=F(x-d)+F(x+d)
\end{equation*}
\begin{equation*}
	\frac{\partial H}{\partial d}=\ln (x+d)+1-\ln (x-d)+(-1)=\ln \frac{x+d}{x-d}>0
\end{equation*}
Hence, the function $H(d)$ increases strictly monotonically. To sum up, we claim that SNE satisfies D1. In addition, Lagrange mean value theorem can also be used to prove the conclusion.

\subsubsection{SNE and D2}\ 

\textit{Theorem B.3}: SNE satisfies D2.
\begin{equation*}
	SNE(\alpha \overset\rightarrow{x})=SNE(\overset\rightarrow{x}),\forall \alpha \in \mathbb{R}, \alpha>0
\end{equation*}

\textit{Proof}:
Suppose $\underbrace{\langle x\rangle}_{N}=A$, so $\underbrace{\langle\alpha x\rangle}_{N}=\alpha A$\\It follows that
\begin{equation*}
	\operatorname{SNE}(\alpha \vec{x})=\frac{\sum_{i=1}^{N}\left(\frac{\alpha x_{i}}{\alpha A} \ln \frac{\alpha x_{i}}{\alpha A}\right)}{N}=\operatorname{SNE}(\vec{x})
\end{equation*}

\subsubsection{SNE and D3}\

\textit{Theorem B.4}: SNE satisfies D3.
\begin{equation*}
	SNE(\alpha+\overset\rightarrow{x})<SNE(\overset\rightarrow{x}),\forall \alpha \in \mathbb{R}, \alpha>0
\end{equation*}

\textit{Proof}: Set 
\begin{equation*}
	F(\alpha)=\sum_{i=1}^{N} \frac{x_{i}+\alpha}{\langle x\rangle+\alpha} \ln \frac{x_{i}+\alpha}{\langle x\rangle+\alpha}
\end{equation*}
It follows that
\begin{equation*}
	\frac{\partial F}{\partial \alpha}=\sum_{i=1}^{N} \frac{\bar{x}-x_{i}}{((\bar{x})+\alpha)^{2}} \ln \frac{x_{i}+\alpha}{\bar{x}+\alpha}+\sum_{i=1}^{N} \frac{\bar{x}-x_{i}}{((\bar{x})+\alpha)^{2}}
\end{equation*}
Clearly, $\sum_{i=1}^{N} \frac{\bar{x}-x_{i}}{((\bar{x})+\alpha)^{2}}=0$; when $\bar{x}\le{x_i}$, $\frac{\bar{x}-x_{i}}{((\bar{x})+\alpha)^{2}} \ln \frac{x_{i}+\alpha}{\bar{x}+\alpha}\le{0}$; and when $\bar{x}>{x_i}$, $\frac{\bar{x}-x_{i}}{((\bar{x})+\alpha)^{2}} \ln \frac{x_{i}+\alpha}{\bar{x}+\alpha}<{0}$. Hence $\frac{\partial F}{\partial \alpha}<{0}$, $F(\alpha)$ is strictly monotonically decreasing. We claim that SNE satisfies D3.
\subsubsection{SNE and D4}\ 

\textit{Theorem B.5}: SNE satisfies D4.
\begin{equation*}
	SNE(\vec{x})=SNE(\vec{x} \| \vec{x})=SNE(\vec{x}\|\vec{x}\| \vec{x})=SNE(\vec{x}\|\vec{x}\| \cdots \| \vec{x})
\end{equation*}

\textit{Proof}:
\begin{equation*}
	\underbrace{\vec{x}\|\vec{x} \cdots\| \vec{x}}_{m}=\underbrace{\langle x\rangle}_{mN}=\underbrace{\langle x\rangle}_{N}=A
\end{equation*}
\begin{equation*}
	SNE(\underbrace{\vec{x}\|\vec{x} \cdots\| \vec{x}}_{m})=\frac{\sum_{i=1}^{N}\left(\frac{m x_{i}}{\underbrace{\langle x\rangle}_{mN}} \ln \frac{m x_{i}}{\underbrace{\langle x\rangle}_{mN}}\right)}{m N}=SNE(\vec{x}) .
\end{equation*}

\subsubsection{SNE and P1}\

\textit{Theorem B.6}: SNE does not satisfy P1.

$\forall i$, $\exists \beta=\beta_{i}>0$, such that $\forall \alpha>0$
\begin{equation*}
	SNE \left ( \left [x_1 \dots x_i+\beta+\alpha \dots  \right ] \right )>SNE\left ( \left [x_1 \dots x_i+\beta \dots \right ] \right )
\end{equation*}

\textit{Proof}: Suppose $N=2$, $\vec{x}=\left[x_{1}, x_{2}\right]$.
We can set
\begin{equation*}
	F(\beta)=\frac{x_{1}}{\langle x\rangle+\beta / 2} \ln \frac{x_{1}}{\langle x\rangle+\beta / 2}+\frac{x_{2}+\beta}{\langle x\rangle+\beta / 2} \ln \frac{x_{2}+\beta}{\langle x\rangle+\beta / 2}
\end{equation*}
It follows that
\begin{equation*}
	\begin{aligned}
		\frac{\partial F}{\partial \beta}&=-\frac{2 x_{1}}{(2 \bar{x}+\beta)^{2}} \ln \frac{2 x_{1}}{2 \bar{x}+\beta}-\frac{2 x_{1}}{(2 \bar{x}+\beta)^{2}}\\&+\frac{4 \bar{x}-2 x_{2}}{(2 \bar{x}+\beta)^{2}} \ln \frac{2 x_{2}+2 \beta}{2 \bar{x}+\beta}+\frac{4 \bar{x}-2 x_{2}}{(2 \bar{x}+\beta)^{2}} \\
		&=\frac{4 \bar{x}-2 x_{2}}{(2 \bar{x}+\beta)^{2}} \ln \frac{2 x_{2}+2 \beta}{2 \bar{x}+\beta}-\frac{2 x_{1}}{(2 \bar{x}+\beta)^{2}} \ln \frac{2 x_{1}}{2 \bar{x}+\beta}\\&=\frac{2 x_{1}}{(2 \bar{x}+\beta)^{2}} \ln \frac{x_{2}+\beta}{x_{1}}
	\end{aligned}
\end{equation*}
For $\forall i$, $\frac{\partial F}{\partial \beta}$ could be negative. Hence SNE does not satisfy P1.

\subsubsection{SNE and P2}\ 

\textit{Theorem B.7}: SNE satisfies P2.
\begin{equation*}
	SNE(\vec{x} \| 0)>SNE(\vec{x})
\end{equation*}

\textit{Proof}:
\begin{equation*}
	\underbrace{\langle x \| 0\rangle}_{N+1}=\frac{N A}{N+1}
\end{equation*}
\begin{equation*}
	\begin{aligned}
		SNE(\vec{x} \| 0)&=\frac{\sum_{i=1}^{N+1}\left(\frac{(N+1) x_{i}}{N A} \ln \frac{(N+1) x_{i}}{N A}\right)}{N+1}\\&=\frac{\sum_{i=1}^{N}\left(\frac{x_{i}}{A} \ln \frac{(N+1) x_{i}}{N A}\right)}{N}
	\end{aligned}
\end{equation*}
Clearly, $\frac{x_{i}}{A} \ln \frac{(N+1) x_{i}}{N A}>\frac{x_{i}}{A} \ln \frac{x_{i}}{A}$. Hence, SNE satisfies P2.

\section*{Acknowledgment}

The authors would like to thank...

\ifCLASSOPTIONcaptionsoff
  \newpage
\fi



%
\bibliographystyle{IEEEtran}
\bibliography{bib}

\end{document}